\documentclass[12pt]{article}
\usepackage{graphicx}
\usepackage{latexsym,amsmath,amsfonts,amssymb}
\newcommand{\be}{\begin{equation}}
\newcommand{\ee}{\end{equation}}
\newcommand{\beq}{\begin{equation}}
\newcommand{\eeq}{\end{equation}}
\newcommand{\bea}{\begin{eqnarray}}
\newcommand{\eea}{\end{eqnarray}}

\newcommand{\ba}{\begin{eqnarray}}
\newcommand{\ea}{\end{eqnarray}}
\begin{document}
\baselineskip=15.5pt
\pagestyle{plain}
\setcounter{page}{1}
%--------+---------+---------+---------+---------+---------+---------+
%Body

% Ofer's definitions

\def\del{{\partial}}
\def\vev#1{\left\langle #1 \right\rangle}
\def\cn{{\cal N}}
\def\co{{\cal O}}
%\newfont{\Bbb}{msbm10 scaled 1200}     %instead of eusb10
%\newcommand{\mathbb}[1]{\mbox{\Bbb #1}}
\def\IC{{\mathbb C}}
\def\IR{{\mathbb R}}
\def\IZ{{\mathbb Z}}
\def\RP{{\bf RP}}
\def\CP{{\bf CP}}
\def\Poincare{{Poincar\'e }}
\def\tr{{\rm tr}}
\def\tp{{\tilde \Phi}}

\def\TL{\hfil$\displaystyle{##}$}
\def\TR{$\displaystyle{{}##}$\hfil}
\def\TC{\hfil$\displaystyle{##}$\hfil}
\def\TT{\hbox{##}}
\def\HLINE{\noalign{\vskip1\jot}\hline\noalign{\vskip1\jot}}
\def\seqalign#1#2{\vcenter{\openup1\jot
  \halign{\strut #1\cr #2 \cr}}}
\def\lbldef#1#2{\expandafter\gdef\csname #1\endcsname {#2}}
\def\eqn#1#2{\lbldef{#1}{(\ref{#1})}%
\begin{equation} #2 \label{#1} \end{equation}}
\def\eqalign#1{\vcenter{\openup1\jot
    \halign{\strut\span\TL & \span\TR\cr #1 \cr
   }}}
\def\eno#1{(\ref{#1})}
\def\href#1#2{#2}
\def\half{{1 \over 2}}

%--------+---------+---------+---------+---------+---------+---------+
%Hirosi's macros:
\def\ads{{\it AdS}}
\def\adsp{{\it AdS}$_{p+2}$}
\def\cft{{\it CFT}}

\newcommand{\ber}{\begin{eqnarray}}
\newcommand{\eer}{\end{eqnarray}}

\newcommand{\beqar}{\begin{eqnarray}}
\newcommand{\cN}{{\cal N}}
\newcommand{\cO}{{\cal O}}
\newcommand{\cA}{{\cal A}}
\newcommand{\cT}{{\cal T}}
\newcommand{\cF}{{\cal F}}
\newcommand{\cC}{{\cal C}}
\newcommand{\cR}{{\cal R}}
\newcommand{\cW}{{\cal W}}
\newcommand{\eeqar}{\end{eqnarray}}
\newcommand{\tht}{\thteta}
\newcommand{\lm}{\lambda}\newcommand{\Lm}{\Lambda}
\newcommand{\eps}{\epsilon}

%--------+---------+---------+---------+---------+---------+---------+

\newcommand{\nonu}{\nonumber}
\newcommand{\oh}{\displaystyle{\frac{1}{2}}}
\newcommand{\dsl}
  {\kern.06em\hbox{\raise.15ex\hbox{$/$}\kern-.56em\hbox{$\partial$}}}
\newcommand{\id}{i\!\!\not\!\partial}
\newcommand{\as}{\not\!\! A}
\newcommand{\ps}{\not\! p}
\newcommand{\ks}{\not\! k}
\newcommand{\D}{{\cal{D}}}
\newcommand{\dv}{d^2x}
\newcommand{\Z}{{\cal Z}}
\newcommand{\N}{{\cal N}}
\newcommand{\Dsl}{\not\!\! D}
\newcommand{\Bsl}{\not\!\! B}
\newcommand{\Psl}{\not\!\! P}
\newcommand{\eeqarr}{\end{eqnarray}}
\newcommand{\ZZ}{{\rm \kern 0.275em Z \kern -0.92em Z}\;}
%--------------------------------Alfonso's definitions%%%%%%%%%%%%%

% DEFINITIONS
                                                                                                    
\def\del{{\delta^{\hbox{\sevenrm B}}}} \def\ex{{\hbox{\rm e}}}
\def\azb{A_{\bar z}} \def\az{A_z} \def\bzb{B_{\bar z}} \def\bz{B_z}
\def\czb{C_{\bar z}} \def\cz{C_z} \def\dzb{D_{\bar z}} \def\dz{D_z}
\def\im{{\hbox{\rm Im}}} \def\mod{{\hbox{\rm mod}}} \def\tr{{\hbox{\rm Tr}}}
\def\ch{{\hbox{\rm ch}}} \def\imp{{\hbox{\sevenrm Im}}}
\def\trp{{\hbox{\sevenrm Tr}}} \def\vol{{\hbox{\rm Vol}}}
\def\rl{\Lambda_{\hbox{\sevenrm R}}} \def\wl{\Lambda_{\hbox{\sevenrm W}}}
\def\fc{{\cal F}_{k+\cox}} \def\vev{vacuum expectation value}
\def\nodiv{\mid{\hbox{\hskip-7.8pt/}}}
\def\ie{{\em i.e.}}
\def\ie{\hbox{\it i.e.}}

\def\CC{{\mathchoice
{\rm C\mkern-8mu\vrule height1.45ex depth-.05ex
width.05em\mkern9mu\kern-.05em}
{\rm C\mkern-8mu\vrule height1.45ex depth-.05ex
width.05em\mkern9mu\kern-.05em}
{\rm C\mkern-8mu\vrule height1ex depth-.07ex
width.035em\mkern9mu\kern-.035em}
{\rm C\mkern-8mu\vrule height.65ex depth-.1ex
width.025em\mkern8mu\kern-.025em}}}
                                                                                                    
\def\RR{{\rm I\kern-1.6pt {\rm R}}}
\def\NN{{\rm I\!N}}
\def\ZZ{{\rm Z}\kern-3.8pt {\rm Z} \kern2pt}
\def\IB{\relax{\rm I\kern-.18em B}}
\def\ID{\relax{\rm I\kern-.18em D}}
\def\II{\relax{\rm I\kern-.18em I}}
\def\IP{\relax{\rm I\kern-.18em P}}
\newcommand{\CS}{{\scriptstyle {\rm CS}}}
\newcommand{\CSs}{{\scriptscriptstyle {\rm CS}}}
\newcommand{\rc}{\nonumber\\}
\newcommand{\bear}{\begin{eqnarray}}
\newcommand{\eear}{\end{eqnarray}}
\newcommand{\W}{{\cal W}}
\newcommand{\F}{{\cal F}}
\newcommand{\x}{{\cal O}}
\newcommand{\LL}{{\cal L}}
                                                                                                    
\def\mani{{\cal M}}
\def\calo{{\cal O}}
\def\calb{{\cal B}}
\def\calw{{\cal W}}
\def\calz{{\cal Z}}
\def\cald{{\cal D}}
\def\calc{{\cal C}}
\def\to{\rightarrow}
\def\ele{{\hbox{\sevenrm L}}}
\def\ere{{\hbox{\sevenrm R}}}
\def\zb{{\bar z}}
\def\wb{{\bar w}}
\def\nodiv{\mid{\hbox{\hskip-7.8pt/}}}
\def\menos{\hbox{\hskip-2.9pt}}
\def\dr{\dot R_}
\def\drr{\dot r_}
\def\ds{\dot s_}
\def\da{\dot A_}
\def\dga{\dot \gamma_}
\def\ga{\gamma_}
\def\dal{\dot\alpha_}
\def\al{\alpha_}
\def\cl{{closed}}
\def\cls{{closing}}
\def\vev{vacuum expectation value}
\def\tr{{\rm Tr}}
\def\to{\rightarrow}
\def\too{\longrightarrow}

% Umut likes:

\def\a{\alpha}
\def\b{\beta}
\def\c{\gamma}
\def\d{\delta}
\def\e{\epsilon}           % Also, \varepsilon
\def\f{\phi}               %      \varphi
\def\vf{\varphi}  \def\tvf{\tilde{\varphi}}
\def\vp{\varphi}
\def\g{\gamma}
\def\h{\eta}
\def\i{\iota}
\def\j{\psi}
\def\k{\kappa}                    % Also, \varkappa (see below)
\def\l{\lambda}
\def\m{\mu}
\def\n{\nu}
\def\o{\omega}  \def\w{\omega}
\def\q{\theta}  \def\th{\theta}                  %     \vartheta
\def\r{\rho}                                     %     \varrho
\def\s{\sigma}                                   %     \varsigma
\def\t{\tau}
\def\u{\upsilon}
\def\x{\xi}
\def\z{\zeta}
\def\pt{\tilde{\varphi}}
\def\tt{\tilde{\theta}}
\def\lab{\label}  
\def\6{\partial}
\def\wg{\wedge}
\def\atanh{{\rm arctanh}}
\def\bpsi{\bar{\psi}}
\def\bt{\bar{\theta}}
\def\bvf{\bar{\varphi}}

%
% FONTS
                                                                                                    
%\newfont{\headfont}{cmbx10 scaled 1440}
\newfont{\namefont}{cmr10}
%\newfont{\initialfont}{cmr10 scaled 1200}
\newfont{\addfont}{cmti7 scaled 1440}
\newfont{\boldmathfont}{cmbx10}
%\newfont{\figfont}{cmr7 scaled 1200}
\newfont{\headfontb}{cmbx10 scaled 1728}
\renewcommand{\theequation}{{\rm\thesection.\arabic{equation}}}
\begin{titlepage}

\begin{center} \Large \bf Elaborations on the String Dual to ${\cal N}=1$ 
SQCD

\end{center}

\vskip 0.3truein
\begin{center}
Roberto Casero${}^{*}$\footnote{roberto.casero@cpht.polytechnique.fr}, 
Carlos 
N\'u\~nez${}^{\dagger}$\footnote{c.nunez@swansea.ac.uk} and Angel 
Paredes${}^{*}$\footnote{angel.paredes@cpht.polytechnique.fr}
\vspace{0.7in}\\
${}^{*}$ \it{Centre de Physique Th\'eorique\\ \'Ecole Polytechnique\\
and UMR du CNRS 7644\\
91128 Palaiseau, France}
\vspace{0.3in}
\vskip 0.2truein
${}^{\dagger}$ \it{Department of Physics\\ University of Swansea, Singleton 
Park\\
Swansea SA2 8PP\\ United Kingdom.}
\vspace{0.3in}
\end{center}
\vskip 1cm
%\begin{center}
\centerline{{\bf Abstract}}

In this paper we make further refinements to the duality proposed between 
$N=1$ SQCD and certain string (supergravity plus branes) backgrounds, 
working in the regime of comparable large number of colors and flavors. 
Using the string theory solutions, we  
predict different field theory observables and phenomena 
like Seiberg duality,  gauge 
coupling and its running, the behavior of Wilson and 't Hooft loops, 
anomalous dimensions of the quark superfields, quartic superpotential 
coupling and its running, continuous and discrete anomaly matching. 
We also give evidence for the smooth interpolation between higgsed and 
confining vacua. We provide several
 matchings between field theory and string theory computations.

%\end{center}

\vskip1truecm
\vspace{0.1in}
\leftline{CPHT-RR 143.0907}
%\leftline{SWAT}
%\leftline{hep-th/0602027 }
\smallskip
\end{titlepage}
\setcounter{footnote}{0}
\tableofcontents
%--------+---------+---------+---------+---------+---------+---------+
%Body
%\Tableofcontents
\section{Introduction}
Maldacena proposed the AdS/CFT Conjecture \cite{Maldacena:1997re} almost 
ten years ago and since 
then, great advances have taken place to 
refine and extend the original duality (see 
\cite{Gubser:1998bc},\cite{Itzhaki:1998dd} and references to those 
articles). 
One of the lines of research that captured the interest of many physicists 
is the use of the ideas and  framework presented in 
\cite{Maldacena:1997re}-\cite{Itzhaki:1998dd}, to formulate duals to 
phenomenologically relevant field theories. The final aim of this line is 
to find a dual to large $N_c$ QCD, that according to early ideas of t'Hooft 
\cite{'t Hooft:1973jz} should 
be described by a theory of strings. 

A slightly less ambitious project (and perhaps a necessary step to 
overtake the case of QCD) is to find duals to 
phenomenologically viable field theories with minimal SUSY. 
Again, remarkable advances on this area of research have been achieved in 
the last nine years, see for example \cite{Girardello:1999bd} and 
references to those works. One well-known limitation of the papers 
mentioned in  \cite{Girardello:1999bd} and successive work on that line 
is the fact that the dual field theory contains fields transforming 
only in the adjoint representation (also in bi-fundamental 
representations for the cascading theories), 
hence they could be thought of as SUSY versions of 
Yang-Mills. The rich dynamics added by the presence of fields transforming 
non-trivially under the 
center of the gauge group, was absent in the 
examples of \cite{Girardello:1999bd}.

To ameliorate this situation the introduction of `flavor' branes (that is 
spacetime filling branes usually with higher dimensionality than the 
`color' branes) was considered. Indeed, the flavor degrees of freedom need 
the presence of an open string sector, and this can only be 
achieved by placing in the background $N_f$ flavor branes, on which the 
$SU(N_f)$ symmetry will be realized. This problem was first treated in the 
`probe' approximation, where the small number of $N_f$ flavor branes was 
not changing the geometry generated by the large number of $N_c$ color 
branes. This non-backreacting 
approximation has a diagrammatic counterpart in the dual QFT,  called 
the quenched approximation. The basic idea is that when 
fundamentals run on internal loops of the diagram, the putative diagram is 
weighted by powers of $N_f/N_c$ and  therefore suppressed in 
this approximation. In other words, the quenched approximation  
boils down to 
considering the path integral of a theory with adjoints and fundamentals
and approximating it 
by the path integral of the theory with adjoints only, where the 
fundamentals 
are only external states (or run on external lines). The approximation 
is good if the quarks are very massive, hence difficult to pair produce, 
or if $N_f/N_c \to 0$ and this is the non-backreacting condition 
mentioned above. 
The papers that shaped this line of research are 
listed in \cite{Karch:2002sh} and there were interesting 
follow-up works. The outcome is a set of spectra of mesons for different 
QFT's duals, a nice picture of chiral symmetry breaking and phase 
transitions.

A natural question then is if some relevant dynamics is lost when 
working in the t'Hooft expansion $N_c\to \infty,\; N_f=fixed$. 
Indeed, one can see that this is the case. From experience in lattice 
field theory, the quenched approximation is good to compute static 
properties (like the spectrum of QCD), but works  poorly when trying to 
address thermodynamical, phase transitions or finite density problems (see 
for example \cite{Karsch:2001cy}). 
Obviously quenching will be 
a bad approximation if the quarks are light and/or if the number of 
fundamentals is comparable to the number of adjoints. 

A natural problem is then to understand how to backreact with 
the flavor branes or, in
the dual field theory, how to `un-quench' the effects of the 
fundamentals. From the diagrammatic viewpoint, this unquenching appears 
when considering the Veneziano Topological Expansion \cite{Veneziano:1976wm} 
instead of the 
t'Hooft expansion \cite{'t Hooft:1973jz}.

On the string side, this problem did not go unnoticed and the approach 
proposed was to find 
solutions to an action that puts together the supergravity 
action (for the color 
branes) 
and the Born-Infeld-Wess-Zumino action (for the flavor branes).
\beq
S= S_{gravity}+ S_{branes}= S_{II B/A} + S_{BI}+ S_{WZ}
\label{sa1}
\eeq
Cases where the flavor branes dynamics, represented by $S_{branes}$ in 
eq.(\ref{sa1}) are localized by some delta-function (that is, finding 
solutions in pure supergravity for 
intersecting $D_p$-$D_{p+4}$ branes), have been discussed, 
see for example \cite{Burrington:2004id}. 
In these cases, typically there is a curvature singularity at the position of the
delta-function, which may hide some of the interesting dynamics arising from
$S_{branes}$.

On the other hand, cases where $S_{branes}$ has support in the
whole spacetime were first 
studied in  non-critical string set-ups \cite{Klebanov:2004ya}
and later in type II 
supergravity where a variety of examples were constructed, ranging from 
duals to a version of
minimally Supersymmetric QCD (SQCD) \cite{Casero:2006pt} and 
\cite{Casero:2007pz}, $N=2$ 
version of SQCD 
\cite{Paredes:2006wb} and the addition of fundamentals to cascading field 
theories/backgrounds 
\cite{Benini:2006hh}. Recently, a 
localized solution for the $D4-D8-\bar D8$ system was found in 
\cite{Burrington:2007qd}.
Some elaborations on the thermodynamics of
the $N=1$ SQCD example mentioned above can be found in 
\cite{Bertoldi:2007sf} and applications to SUSY breaking in 
\cite{Hirano:2007cj}. From a more formal point of view it was shown in 
\cite{Koerber:2007hd} that our intuitive way of adding flavor branes can 
be put on a formal basis where SUSY is guaranteed whenever the flavor 
branes 
are calibrated, and  the second-order Einstein equations are 
automatically satisfied once the first-order BPS and Maxwell equations 
and Bianchi identities with sources are solved.

Notice that in the context of critical string theory, backreacting with 
the flavor branes as indicated in eq.(\ref{sa1}) is a natural continuation 
of the approach in \cite{Karch:2002sh} and indeed 
the correct way to introduce 
the fundamental degrees of freedom  with $SU(N_f)$ global symmetry.

In this paper, we will study the dual to minimally 
SUSY QCD that was proposed by the present authors in 
\cite{Casero:2006pt}.
Indeed, in \cite{Casero:2006pt} a set of  solutions to type IIB 
supergravity plus 
branes has been found and a large number of checks presented, where
it was shown that the type IIB backgrounds were capturing non-perturbative 
effects of this version of SQCD.

We will present a new set of  solutions to type IIB 
supergravity plus branes, dual to the field theory in different vacua and 
will provide a large number of matchings 
to show that our new solutions indeed capture new qualitative and 
quantitative information of the dual QFT.

The organization of this paper is as follows: in section \ref{seccion2}
we will review some material from~\cite{Casero:2006pt}, set up the 
notation for the string backgrounds and dual SQCD. We also discuss some 
aspects of the SQCD-like theory that can be learned from the background 
without knowing the explicit solution, like Seiberg duality and the matching of 
R-symmetry anomalies, and we discuss some aspects of the field theory that 
will be relevant for the rest of the paper. In section~\ref{sectionsolutions} we 
present the new solutions that motivate this paper. 
We show that an expansion for 
large values of the radial coordinate (the UV of the dual SQCD) can be 
smoothly connected with the IR expansion of the solution. We discuss in 
detail the numerics and present plots to clarify ideas.
In section \ref{predictions} we study the predictions that these new 
solutions make for the physics of the SQCD theory; they include definitions of 
gauge coupling, anomalous dimensions, quartic coupling and computation of the beta 
function, continuous and discrete anomaly matching, potential between  
quark-antiquark and 
monopole-antimonopole pairs  as computed by Wilson and~'t~Hooft loops and 
different aspects of  the QFT 
that we learn from the string solution. We close with some 
conclusions and future work to appear. We complement our presentation with 
an appendix on a nice technical subtlety.
\section{Brief Review of the dual to SQCD}\label{seccion2}
\setcounter{equation}{0}
In this section, we will go over some points of \cite{Casero:2006pt}
that will be relevant in the rest of the paper. First, we will review the
two types of backgrounds that follow from the work \cite{Casero:2006pt}.
After that, we will discuss some aspects of the dual QFT.
\subsection{The two types of backgrounds}
We presented in \cite{Casero:2006pt} a string dual to a version of $N=1$ 
SQCD. 
The construction proceeded as explained around eq.(\ref{sa1}) by adding 
flavor branes to a given supergravity solution and finding a new 
background where the effects of flavor branes were encoded.
The {\it unflavored} solution we based our construction on, was 
proposed to be 
the dual to $N=1$ SYM in \cite{Maldacena:2000yy} (the solution was 
originally found in 4d gauged supergravity in \cite{Chamseddine:1997nm}).
We considered the backreaction of the flavor branes, that in this 
particular case are $N_f$ $D5$ branes wrapping a non-compact two-cycle. 
We found 
two types of solutions, which in the following we will refer to as solutions of 
type {\bf A} and solutions of type {\bf N}.

Backgrounds of type {\bf A} look quite simple, consist on a metric, a 
Ramond 
three-form, a dilaton and they read, in Einstein frame and setting 
$\alpha'= g_s =1$ (we have slightly changed the notation with
respect to  \cite{Casero:2006pt}),
\bea
ds^2& =& e^{\frac{\phi(\rho)}{2}} \Big[ dx_{1,3}^2 + 4 Y(\rho) d\rho^2 + 
H(\rho)
(d\theta^2 + \sin^2\theta d\varphi^2)
+G(\rho) (d\tilde\theta^2 +\sin^2\tilde\theta 
d\tilde\varphi^2)\nonumber\\
& & + Y(\rho) 
(d\psi +\cos\tilde\theta d\tilde\varphi + \cos\theta d\varphi)^2 \Big]  
\nonumber\\
F_{(3)}&=&- \Big[\frac{N_c }{4} \sin\tilde\theta d\tilde\theta \wedge 
d\tilde\varphi 
+\frac{N_f - N_c}{4} 
\sin\theta d\theta \wedge d\varphi \Big] \wedge ( d\psi + \cos 
\tilde \theta d\tilde\varphi +\cos\theta d\varphi),\,\,\nonumber\\
\phi&=&\phi(\rho)\, ,
\label{metricA}
\eea
where we have used the coordinates $x=(x^\mu, 
\rho,\theta,\varphi,\tilde\theta,\tilde\varphi,\psi)$. The 
functions in eq.(\ref{metricA}), satisfy a set of BPS eqs (that solve 
the Einstein, Maxwell and Bianchi eqs). The derivation of these BPS eqs was discussed in detail in 
\cite{Casero:2006pt}. They read:
\ba
H'&=& \frac{1}{2} (N_c-N_f)  + 2 Y
\label{newbps1} \\
G' &=& -\frac{N_c}{2} + 2Y
\label{newbps2} \\
Y' &=&-\frac{1}{2}(N_f- N_c) \frac{Y}{H} -\frac{N_c}{2} \frac{Y}{G} 
-2Y^2(\frac{1}{H}+\frac{1}{G}) +4Y
\label{newbps3}  \\
\phi'&=&-\frac{ (N_c-N_f)}{4H}   + \frac{N_c}{4G} 
\label{newbps4}
\ea
On the other hand, solutions of type {\bf N}£ are slightly more involved 
and can be seen as a generalization of the type {\bf A} backgrounds. The 
dilaton is a function of the radial coordinate $\phi=\phi(\rho)$, while 
the metric and Ramond three form read,
\ba
ds^2 &=& e^{ \frac{\phi(\rho)}{2}} \Big[dx_{1,3}^2 + e^{2k(\rho)}d\rho^2 
+ e^{2 h(\rho)} 
(d\theta^2 + \sin^2\theta d\varphi^2) +\nonumber\\
&+&\frac{e^{2 g(\rho)}}{4} 
\left((\tilde{\omega}_1+a(\rho)d\theta)^2 
+ (\tilde{\omega}_2-a(\rho)\sin\theta d\varphi)^2\right)
 + \frac{e^{2 k(\rho)}}{4} 
(\tilde{\omega}_3 + \cos\theta d\varphi)^2\Big] \nonumber\\
F_{(3)} &=&\frac{N_c}{4}\Bigg[-(\tilde{\omega}_1+b(\rho) d\theta)\wedge
(\tilde{\omega}_2-b(\rho) \sin\theta d\varphi)\wedge
(\tilde{\omega}_3 + \cos\theta d\varphi)+\nonumber\\
& & b'd\rho \wedge (-d\theta \wedge \tilde{\omega}_1  + 
\sin\theta d\varphi 
\wedge 
\tilde{\omega}_2) + (1-b(\rho)^2) \sin\theta d\theta\wedge d\varphi \wedge
\tilde{\omega}_3\Bigg]\nonumber\\
&-&\frac{N_f}{4}\sin\theta d\theta \wedge 
d\varphi \wedge (d\psi +\cos\tilde{\theta} d\tilde{\varphi}).
\label{nonabmetric424}
\ea
Where $\tilde\omega_i$ are the left-invariant forms of $SU(2)$
\bea\lab{su2}
\tilde{\omega}_1&=& \cos\psi d\tilde\theta\,+\,\sin\psi\sin\tilde\theta
d\tilde\varphi\,\,,\rc
\tilde{\omega}_2&=&-\sin\psi d\tilde\theta\,+\,\cos\psi\sin\tilde\theta
d\tilde\varphi\,\,,\rc
\tilde{\omega}_3&=&d\psi\,+\,\cos\tilde\theta d\tilde\varphi\,\,.
\eea
For the case of backgrounds of type {\bf N} we also found a set of BPS 
eqs and solutions to it, by expanding in series near 
$\rho=0$ 
(the IR of the dual QFT) and $\rho\to\infty$ (the UV of the dual QFT) 
and then showing that the two expansions can be numerically matched in a  
smooth way.
It was for the backgrounds of  type {\bf N}  that we 
presented a considerable number of checks showing 
how they captured non-perturbative aspects of 
SQCD \cite{Casero:2006pt}.

In this paper, we will focus on backgrounds of type {\bf A} above, finding 
a new set of solutions for the functions $H,G,Y,\phi$ and  we will 
study the strong-coupling effects that they predict for the dual QFT. This kind of solutions was first
overlooked in \cite{Casero:2006pt}, because at that time we could only find badly singular type {\bf A} 
solutions. We believe the new set of backgrounds we derive in the present paper, which are 
characterized by a good IR singularity, nicely complement the 
overall picture.

After having set up the stage, it is now convenient to discuss some 
subtleties 
about the dual theory, that as we anticipated above, is a version of $N=1$ 
SQCD.
\subsection{The proposed dual field theory}
\label{proposed}
As it is well known, all duals to non-conformal gauge field theories relying 
on the supergravity approximation 
like the ones in \cite{Girardello:1999bd}, \cite{Maldacena:2000yy} or any 
other model available,  
are characterized by the presence of extra modes or some  other kind of 
UV completion
\footnote{This can be avoided in the non-critical string 
approach, but in that case the supergravity 
approximation is not reliable.}.
This UV completion is not necessarily an ugly feature, one could perhaps 
be 
interested in the field 
theory with the UV completion that string theory provides (like in the 
case of cascading theories) or perhaps, 
there are ways to distinguish when a given field theory correlator is 
affected or not by the extra modes. In the particular case of interest 
in 
this paper, the field theory 
dual to the background in \cite{Maldacena:2000yy} 
\cite{Chamseddine:1997nm}
was proposed in  
the paper \cite{Maldacena:2000yy} to be $N=1$ SYM plus an infinite set of 
KK modes, which is 
just another way of saying that the UV completion is a higher dimensional 
theory. The dynamics 
of these extra modes was studied in detail in \cite{Andrews:2005cv}
and a way to disentangle the effects of the UV completion from the 
Super-Yang-Mills modes was proposed in \cite{Gursoy:2005cn}. In summary, 
the picture of the dual field theory of \cite{Maldacena:2000yy} is 
clear by now. Simplifying things a bit, one could say that the field 
theory without flavors is described by a lagrangian of the form (in 
fields and superfields notation)
\bea
 L &=& -Tr[\frac{1}{4} F_{\mu\nu}^2 + i\lambda D\lambda- \sum_k 
(D_\mu\phi_k)^2 - M_{KK}^2 \phi_k^2
+ \psi_k (i D - M_{KK})\psi_k
+ V(\phi_k, \psi_k,\lambda)]=\nonumber\\
 & =&\int d^4\theta \sum_k \Phi^\dagger_k e^V \Phi_k +\int d^2\theta 
\Big(W_\alpha W^\alpha +\sum_k {\cal W}(\Phi_k)\Big) + h.c.
\label{lagrangiantwisted}
\eea
where, schematically, we wrote a lagrangian describing $N=1$ SYM coupled to 
an infinite tower of SUSY KK modes, with a given superpotential 
${\cal{W}}(\Phi_k)$ (for a detailed expression see~\cite{Andrews:2005cv}).

We proposed in \cite{Casero:2006pt} that the addition of the flavor branes 
introduces two chiral superfields $Q,\tilde{Q}$ transforming in the 
fundamental and anti-fundamental of the $SU(N_c)$ gauge group, respectively. 
The dynamics and the
coupling of the quark superfields with the already existing adjoints was 
proposed to be 
of the form,
\beq
L= \int d^4 \theta (Q^\dagger e^V Q + \tilde{Q}^\dagger e^{V} \tilde{Q}) + 
\int d^2\theta \tilde{Q} \Phi_k Q + h.c.
\label{quarks}
\eeq
where $\Phi_k$ is a generic adjoint field. So, putting together 
(\ref{lagrangiantwisted}) and (\ref{quarks}), we have 
a lagrangian that in superspace will schematically read,
\beq
L= \int d^4\theta \Big(  Q^\dagger e^V Q + \tilde{Q}^\dagger e^{V} 
\tilde{Q} 
+\sum_k \Phi^{\dagger}_k e^{V}\Phi_k\Big) 
+ \int d^2\theta \Big(W_\alpha W^\alpha + \sum_k \tilde{Q} \Phi_k Q + 
{\cal W}(\Phi_k) \Big)+ h.c.
\label{lagrangianflavors}
\eeq
Let us consider the simplest 
case in which the superpotential contains only a 
mass term ${\cal W}(\Phi_k)= \mu_k \,\Phi_k^2$. The 
occasional presence of 
higher powers of the chiral multiplet of KK modes will be irrelevant for 
the IR dynamics as we 
will see below. Since we are interested in studying the field theory 
at low energies, we can integrate out the scalar KK modes. This should be 
taken with some care because the mass scale of the KK modes is comparable 
with the scale of strong coupling which we are interested in, so the 
integration-out of these additional modes is not totally clean. After 
integrating over the dynamics 
of 
the scalar multiplet, we get a dynamics for the quarks and vector 
superfields of the form
\beq
L= \int d^4\theta \left( Q^\dagger e^V Q + \tilde{Q}^\dagger e^{V} \tilde{Q}\right)
+ \int d^2\theta  \left(W_\alpha W^\alpha + 
\frac{1}{\mu}\Big[Tr(\tilde{Q}Q)^2-\frac{1}{N_c}(Tr \tilde{Q}Q)^2  \Big]+ 
O(\tilde{Q}Q^3)\right)
\label{ourtheory}
\eeq
So, we arrive to a theory with the field content of $N=1$ SQCD, but with 
an important dynamical difference: the flavor symmetry is $SU(N_f)$. The 
root of this 
difference is basically that the `mother theory' (\ref{lagrangianflavors}) 
has already $SU(N_f)$ and not $SU(N_f)_l \times SU(N_f)_r$ because of the 
term $\tilde{Q} \Phi_k Q $. This same qualitative difference from SQCD 
or QCD will generically appear when adding flavors to  
field theories that could be thought of as coming from an $N=2$ theory. 
On the other hand, when color and flavor branes are  
codimension six, as for instance in the last paper of 
ref.\cite{Karch:2002sh} and 
\cite{Casero:2007ae,Burrington:2007qd},  this kind of coupling is not 
present.
Another interesting point is that if the adjoint superfield transforms in 
$U(N_c)$ instead of $SU(N_c)$, then the double trace term in the 
superpotential is absent.

In the following, only the quartic coupling for the fundamentals will
be considered. In section \ref{sec: beta}, we will argue that higher
order couplings are irrelevant. 
We present here a heuristic complementary argument
which favors the appearance of the quartic coupling and not the higher
ones upon integrating-out the adjoint scalars. The superpotential 
for the adjoint scalars is of the form \cite{Andrews:2005cv},
\beq
{\cal{W}}(\Phi_k)=Tr\Big[ \Phi_1[\Phi_2,\Phi_3] 
+\sum_i \frac{\mu}{2}\Phi_i^2\Big]\,\,.
\label{doreyandrews}
\eeq
The ``democratic" way in which the $N_f$ flavor branes are treated suggests
that the $\tilde Q Q$ matrix is proportional to the identity (notice that
$\tilde Q Q$ here is not a meson operator, gauge indices are not contracted).
When we integrate out the adjoints, and insert back their value in 
${\cal W}(\Phi_k)$,
the higher order contributions involving $\tilde Q Q$ vanish using conmutation 
relations.

We would like to mention at this point that, in order to get our BPS eqs 
and their solutions, we introduced in \cite{Casero:2006pt} a {\it 
smearing procedure}. This has some effect on the superpotential, as 
discussed in that work. We will not insist here on the fact 
that qualitative and quantitative aspects of the localized solution are 
well captured by the smeared solution. For more discussions, we refer the 
interested reader to our forthcoming paper \cite{cnp3}. 

In the paper \cite{Casero:2006pt}, backgrounds of type {\bf N} in 
eq.(\ref{nonabmetric424}) above were 
shown to capture a variety of non-perturbative aspects for a field 
theory like (\ref{ourtheory}). In the rest of this paper, we will show how 
backgrounds of type {\bf A} also encode interesting dynamics of 
(\ref{ourtheory}). We now turn to 
study some aspects of the theory defined in eq. (\ref{ourtheory}) that 
will be relevant for the 
rest of the paper.
\subsubsection{A couple of important subtleties}
\label{nfbignc}
We want to explain briefly 
some nice subtleties that arise when considering the theory 
(\ref{ourtheory}). Let us start with the superpotential in 
(\ref{ourtheory}),
\beq
W_{tree}=\frac{1}{\mu}\Big[Tr(\tilde{Q}Q^2)-\frac{1}{N_c}(Tr \tilde{Q}Q)^2  
\Big]
\label{wtree}
\eeq
It is known that when analyzing the space of vacua for this theory the 
quantum mechanical superpotential differs from the classical one in eq. 
(\ref{wtree}). Indeed, if $N_f<N_c$ an Affleck-Dine-Seiberg superpotential 
\cite{Affleck:1983mk}
is induced 
\beq
W=W_{tree} + (N_c-N_f)\Big(\frac{\Lambda^{3N_c-N_f}}{\det (\tilde{Q}Q)} 
\Big)^{\frac{1}{N_c-N_f}}
\label{afds}
\eeq
which changes qualitatively the dynamics, for example, when looking for 
solutions to the F-term eqs, not allowing the 
(diagonalized) meson field $M=\tilde{Q} Q$ to have zero eigenvalues.

When $N_f \geq N_c$ a different quantum superpotential is induced. 
Actually, for the cases $N_f=N_c$, $N_f=N_c +1$ and $N_f>N_c$, three 
different superpotentials are induced that present different space of 
vacua. Here, we will content ourselves with stating that for the
case of $N_f \geq N_c$, the field theory presents two possible behaviors, 
one in which 
the meson matrix has non-zero eigenvalues and another branch where the 
meson matrix has zero determinant (the meson matrix can always be 
diagonalized). For details, see 
\cite{Carlino:2000ff}, \cite{ABCMN}.

We will explain in section \ref{sectionsolutions} how these two possible 
behaviors are matched by the supergravity solutions of type {\bf A} and 
type {\bf N}.

The second interesting subtlety refers to the large $\rho$ behavior of the 
solutions we found. As we will see in section \ref{sec: revisit}, the 
careful study of 
the UV (large $\rho$ expansion) of our solutions suggests that in the 
dual field theory an operator of dimension six is getting a  VEV. Probably as a 
consequence of this, some well-known or expected behaviour of the IR 
dynamics of the theory (\ref{ourtheory}), see \cite{Strassler:2005qs}, will not be realized by our 
string background, that in turn is dual to the field theory but in a 
particular vacuum. Indeed, in the following we will check that the string
solutions seem to agree with the interpretation of SQCD with quartic
superpotential in the UV (large $\rho$) region but 
 we will observe that in general the expected IR 
fixed point is not attained. Of course, this depends on the precise
vacuum on which the theory is realized.
As a matter of fact, if the theory flows to the IR Seiberg's fixed point, {\it i.e.}, the
quartic coupling flowed to zero, there is an enhancement of the flavor symmetry at the
IR fixed point. It is hard to see how a brane construction of the kind we 
are
using can reflect this fact. Thus, there must be something in the field theory dual 
to the presented solutions preventing the vanishing of this coupling. It would be very interesting to
understand this point better.
In section \ref{backto1970} we will comment on the actual IR behaviour which can be read 
from the geometry.
\subsubsection{Beta functions}
\label{sec: beta}
We know that in an $N=1$ field theory, there are  lots of simplifications 
for the beta functions of different couplings. For the gauge 
coupling, the well known NSVZ-Jones beta function reads
\beq
\beta_g=-\Big(\frac{g^3/(16\pi^2)}{(1-\frac{g^2 N_c}{8\pi^2})}\Big)\Big(3N_c- 
N_f(1-\gamma_Q)\Big)
\label{betag}
\eeq
where $\gamma_Q$ is the anomalous dimension of the quark superfields.
In the following, we will be interested in the wilsonian beta function, 
for which the denominator in (\ref{betag}) is absent.
Suppose that the superpotential for our model is (here we  neglect the 
double trace terms of the form $\frac{1}{N_c}(Tr\tilde{Q}Q)^2$, that will 
not change the beta functions)
\beq
W=\kappa (Q\tilde{Q})^2 + \lambda (Q\tilde{Q})^3 + z (Q\tilde{Q})^4+...
\label{beta1}
\eeq
Let us remind the standard result in four-dimensional SUSY gauge theories; 
for a 
superpotential of 
the form
$W= h Q^n A^m $ if the fields have (classical) dimensions 
$
[A]= mass^a,\; [Q]= mass^q$
we can form the quantity $\Delta_h= 3 - n q - m a$ and the adimensional 
coupling $\tilde{h}= h \mu^{-\Delta_h}$ ( where $\mu$ is some mass 
scale).
Then, the beta function for the coupling $\tilde{h}$ is
\beq
\beta_{\tilde h}=\tilde{h}[-\Delta_h + \frac{n}{2}\gamma_Q+ 
\frac{m}{2}\gamma_A]
\eeq
where $\gamma_A, \gamma_Q$ are the anomalous dimensions of each field.
Let us apply this to our case (\ref{beta1}),
\beq
\beta_{\tilde \kappa}=\tilde{\kappa}[1 +\frac{4}{2}\gamma_Q  ],\;\;
\beta_{\tilde \lambda}=\tilde{\lambda}[ 3+\frac{6}{2}\gamma_Q  ],\;\;
\beta_{\tilde z}=\tilde{z}[5+ \frac{8}{2}\gamma_Q  ]
\label{betas}
\eeq
We will argue in section \ref{betaanom} that in the solutions
for $N_f < 2N_c$, near $\rho \to \infty$, one finds:
\beq
\gamma_Q=-\frac{1}{2} - \epsilon\,,\qquad (N_f < 2 N_c)
\label{gammauv}
\eeq
with $\epsilon$  some small correction in the UV (but at lower energies 
it may become large). 
Thus, when $N_f<2 N_c$ the 
quartic coupling $\tilde \kappa$ is relevant (becomes marginal
at $N_f = 2N_c$), but the others are 
irrelevant.
When $N_f>2N_c$, we have $\gamma_Q > -\frac12$ and all couplings are 
irrelevant (their beta functions are all positive). 
So we see that dropping $(\tilde{Q}Q)^3$ and higher orders 
from the lagrangian (\ref{ourtheory}) is well motivated if we are 
interested in the IR dynamics. Also, the argument around 
eq.(\ref{doreyandrews}) suggests that the higher order operators do 
not occur at all, so in the following we will consider the case of quartic 
superpotential only. In section \ref{betaanom}, we will show how solutions 
of type {\bf A} reproduce
the beta functions for the gauge coupling in (\ref{betag}) and the 
anomalous dimensions in (\ref{gammauv}). 

Let us now focus on another interesting property of the field theory 
(\ref{ourtheory}), Seiberg duality.
\subsection{Seiberg Duality}
A well known 
property of $N=1$ SQCD-like theories is 
Seiberg duality \cite{Seiberg:1994pq}. In the particular case of field 
theories like the one in eq.(\ref{ourtheory}), this equivalence is 
specially nice as explained for example in \cite{Strassler:2005qs},
where it was shown that the duality can be exact and not just an IR 
equivalence of 
two theories. In \cite{Casero:2006pt}, we explained how the duality 
proposed by Seiberg is manifest in our backgrounds of type {\bf N}. Here, 
we will concentrate on showing how type {\bf A} backgrounds reflect 
this duality.

Indeed, without knowing the solution to the BPS system of 
equations (\ref{newbps1})-(\ref{newbps4}),
we can observe that the equations and hence 
their solutions are {\it invariant} under:
\be
 H\leftrightarrow G,\;\;\;\;\; N_c\to N_f - N_c 
\label{seiberg}
\ee
with $\phi,\, Y,\,N_f$ unchanged.
This change has the correct numerology to act as Seiberg duality. Indeed, 
we can see that the transformation (\ref{seiberg}) is idempotent. 

Let us comment on the correspondence between Seiberg 
duality and the transformation (\ref{seiberg}). The reader may want to 
think about it in this way: if we take backgrounds of the form 
(\ref{metricA}), characterized by BPS eqs.(\ref{newbps1})-(\ref{newbps4}), 
then it is observed that a solution for the functions $H,G, 
Y,\phi$, dual to a field theory with $N_f$ flavors and $N_c$ colors, 
automatically  produces `another' solution for the functions 
$G,H,Y,\phi$ that we can associate with the field theory with $N_f$ 
flavors and $N_f-N_c$ colors. The fact that the two solutions are the same 
solution (up to a coordinate redefinition), indicates that the two 
field theories 
are actually the same theory. This nicely matches with the exactness of 
Seiberg duality for theories like (\ref{ourtheory}), as discussed in 
detail in \cite{Strassler:2005qs}.

One natural question that may appear at this point is whether the Seiberg 
duality change (\ref{seiberg}) is only a property of the SUSY system or, 
on the contrary, all solutions (SUSY and non-SUSY) to the eqs of motion 
that dictate the dynamics of type {\bf A} backgrounds have the property 
above. 
To sort this out, notice that not only the BPS equations, but also
the type {\bf A} ansatz (\ref{metricA}) is invariant under
(\ref{seiberg}) if we relabel:
\be
\theta \leftrightarrow \tilde\theta\,, \qquad
\varphi \leftrightarrow \tilde\varphi
\label{cambi}
\ee
We can see that Seiberg duality 
is basically the interchange of the two-manifolds labelled by 
$(\theta,\varphi)$ and 
$(\tilde\theta,\tilde\varphi)$.
Thus, even non-supersymmetric solutions of the resulting
equations of motion, would have a Seiberg-dual counterpart.

 This kind of 
reasoning should nevertheless be taken carefully. Indeed, Seiberg proved 
\cite{Seiberg:1994pq} that in 
dual pairs there was matching of global anomalies and that 
deformations of moduli spaces could be put in correspondence. It 
is therefore necessary to study how anomalies for the 
R-symmetry are manifest in 
each theory in the dual pair. 
The prescription to compute R-symmetry anomalies was 
studied in \cite{Casero:2006pt} and, in following sections,
 we just apply that information for the 
matching of R-symmetry anomalies on both dual pairs.

Summarizing, we propose the following interpretation: 
since a gravity solution corresponds to a field theory in a given vacuum,
the fact that two different field theories correspond to the
same gravity solution means that they are the same theory,
they contain the same information (they are Seiberg dual).
However, the way of extracting gauge theory information such as couplings
or rank of the gauge group
from the geometry can depend on the interpretation, on how one
wants to understand -in view of the change (\ref{cambi})-
the role of the different cycles in the geometry among other things. 
\subsubsection{R-symmetry anomalies}\label{anomalymatching}
As usual for anomalies, we look into the Ramond fields, in this case 
$C_2$.  
One might be confused, since we write a potential $C_2$ when  
eq. (\ref{metricA}) implies that 
$d F_3 
\neq 0$
(that is the Bianchi identity 
indicates the presence of sources).
The point that we made in 
\cite{Casero:2006pt} was that one should first restrict the 
background to the cycle, 
\beq
\Sigma=\Big(\theta=\tilde{\theta},\;\;\; \varphi= 2\pi -\tilde \varphi, 
\;\;\; \psi\Big),
\label{ciclot}
\eeq
since this is where color branes are wrapped around. On this sub-space $F_3$ is exact and one can write a potential $C_2$. For 
the  theory with  $N_f<2N_c$, that we call {\bf e}-theory (for electric), 
and for the Seiberg 
dual  theory, 
with $N_f>2 n_c$, that we call the {\bf m}-theory (for 
magnetic)\footnote{The relation 
is the 
usual $n_c=N_f-N_c$.}, the potentials are,
\bea
& &C_{2e}=(\psi-\psi_0)\frac{N_c}{4}(2-x)sin\theta d\theta \wedge 
d\varphi\nonumber\\ 
& &C_{2m}=(\psi-\psi_0)\frac{n_c}{4}(\bar{x}-2)sin\tilde{\theta} 
d\tilde{\theta} 
\wedge 
d\tilde{\varphi}
\label{c2}
\eea
We have defined $x=N_f/N_c$ and $\bar{x}=N_f/n_c$ as in 
\cite{Casero:2006pt}. Notice that both expressions in (\ref{c2})
are the same expressed in electric and magnetic variables respectively.

Imposing that the instanton that is represented by a Euclidean 
D1-brane wrapping the contractible two-cycle (\ref{ciclot})
is not observed in Bohm-Aharonov experiment \cite{Gursoy:2003hf}, 
leads to some particular values of the $\psi$ angle (or selects some 
possible translations in $\psi$, if we make $\psi\to \psi+2\epsilon$). 
Indeed, remember that
$\psi\to \psi+ 2\epsilon$ (with $\epsilon$ in $ [0,2\pi]$) 
was the change of R-symmetry in the unflavored case 
\cite{Maldacena:2000yy}
and the unobservability 
of the shrinking instanton implies for the 
partition function of the putative Euclidean brane,
\beq
Z= Z[0] e^{\frac{i}{2\pi}\int C_2}\to 
e^{i\epsilon N_c (2-x) } e^{\frac{i}{2\pi}\int C_2}\times Z[0]=Z
\eeq
where $Z[0]$ is the partition function for this probe including the 
Born-Infeld part; this in turn leads to,
\beq
\epsilon_e= \frac{2\pi n}{2 N_c- N_f},\;\; \epsilon_m= \frac{2\pi k}{ N_f 
-2 n_c}
\label{match}
\eeq
The relation in eq.(\ref{match}) should be interpreted as the matching of 
the anomaly in the correlator of two gauge currents and the R-symmetry 
current, 
$<J(R)J(N_c)J(N_c)>$ on the theories of the dual pair. 
We can see that $\epsilon_e= \epsilon_m$ as $k=(1,....N_f-2n_c)$ and 
$n=(1,...2N_c-N_f)$.
This will work in the same way for the type {\bf N} backgrounds, since it 
is the same RR field $C_2$ that is relevant.
This is {\it not} one of the 
diagrams alluded by the 't Hooft criteria, those will be discussed in 
section~\ref{sec:globmatch}. 

The fact that the R-symmetry is broken by 
anomalies to the discrete subgroup $Z_{2N_c- N_f}$ suggest that the 
R-charge for the quark superfield is $R[Q]=\frac{1}{2}$.  
Indeed, as we will see in section \ref{betaanom} and in section 
\ref{sec:globmatch}  
the result of 
eq.(\ref{match}) implies that the R-charge of the quark superfield is
$R[Q]=1/2$ (in the UV). The anomalous dimension of the 
quark superfields must then be (in the case of an UV fixed point) 
$\gamma_Q=-\frac{1}{2}$, in agreement with what
we observed around eq.(\ref{gammauv}).

\subsection{A jewel of the 1970's}\label{jewel1970}
Before leaving this field theory section, we would like to remind the 
readers about 
nice work done three decades ago.

Several different suggestions have been made to address the problem of confinement. The idea of 't Hooft 
and Mandelstam \cite{'tHooft:1977hy}
was to think about confinement as a dual Meissner 
effect, 
where a `vortex' of chromoelectric flux joined together two quarks giving 
an area law for the Wilson loop.\footnote{Needless to say, this regards 
the problem of confinement on Yang-Mills; in QCD, the presence of 
fundamental matter allows to interpolate between the Higgs and the 
confining phase.}
The `abelian-projection monopoles' proposed to be dual to the quarks
\cite{'tHooft:1982ns} paved the way to 
interpret the phenomena of confinement in QCD as a dual Meissner effect. 
The following possible phases manifest in a non-abelian gauge theory:
\begin{itemize}
\item  Free phase: in analogy with QED; for example, if the number of 
flavors is very large in QCD.

\item  Coulomb phase: the IR theory is conformal.

\item Higgs phase: here the monopoles are confined and quarks are free.

\item Wilson/Confining phase: here the quarks are confined and 
monopoles are free.

\item Oblique confined phase: quarks and monopoles are confined, dyons 
condense.
\end{itemize}
In QCD, SQCD or in any QCD-like theory with fundamental 
matter,  
one can move smoothly (without phase transitions) between the different  
phases mentioned above~\cite{Fradkin:1978dv}.
We will see how the solutions of type {\bf A} that we will find in section 
\ref{sectionsolutions} 
realize these phases and their smooth interpolation.

Let us now present the new solutions for the backgrounds
of type {\bf A} in eqs.(\ref{metricA})-(\ref{newbps4}).

\section{New solutions}
\label{sectionsolutions}
\setcounter{equation}{0}

In this section we present the new solutions for backgrounds of type {\bf 
A} described in eq.(\ref{metricA}). 

Let us first explain why we are interested in 
solutions of type {\bf A}. It is well known that even when singular,
certain supergravity solutions can faithfully describe 
the dynamics of a dual gauge theory at strong coupling.
The question of {\it which} singularities are allowed inspired a variety 
of 
criteria to discard those singular spacetimes that cannot be proposed 
to be dual to a given gauge theory and accept those that can. Among these 
criteria, we will choose 
the one in \cite{Maldacena:2000mw}, that basically imposes on a singular 
spacetime to have finite value of $g_{tt}$ at the singularity
(in this discussion
 we are alluding to IR singularities, the UV singularities of these
kind of theories are resolved in a stringy way).

Different studies with supergravity solutions dual to field theories 
without flavors, have exemplified the correctness of this criterion. It is 
common lore by now (in the case of `unflavored' backgrounds), that 
solutions of the type {\bf A} are singular with a so called ``bad 
singularity'', that is, the putative background cannot be interpreted as being
dual to a field theory, see for example \cite{Maldacena:2000yy}. 

In our case, backgrounds of type {\bf A} will have associated 
a value of $g_{tt}= e^{\frac{\phi}{2}}$ (in Einstein frame). If we analyze 
the BPS 
eq. for the dilaton- see the last eq of (\ref{newbps4})- we immediately 
see 
that the dilaton has positive derivative if $N_f \geq N_c$. 
Hence, type {\bf A} backgrounds present a ``good
singularity'' when the number of flavors is bigger or equal to the number 
of colors. This restriction very nicely matches what we explained 
in section \ref{nfbignc} regarding the two possible vacua for $N_f \geq 
N_c$. In other words \footnote{This interpretation was developed in 
discussions with Francesco Bigazzi and Aldo Cotrone.}, backgrounds of 
type {\bf A} describe 
N=1 SQCD with the lagrangian (\ref{ourtheory}) when the gaugino 
condensate and the VEVs of the meson 
matrix vanish. Conversely, we propose that backgrounds of type {\bf N} 
studied in detail in \cite{Casero:2006pt} describe the same field theory, 
in a vacuum with non-vanishing meson VEV and gaugino 
condensate (we remind the reader that, in the unflavored case,
it is well established that the function $a(\rho)$ 
in (\ref{nonabmetric424}) is,
roughly, the dual of the gaugino condensate and that type {\bf A}
solutions are obtained from type {\bf N} by seeting $a(\rho)=0$).

In summary, for $N_f \geq N_c$, type {\bf A} backgrounds can be 
thought of as good duals to our $N=1$ SQCD of eq.(\ref{ourtheory}) in a 
vacuum with vanishing meson VEV and gaugino condensate. To describe 
these new solutions, we now present in detail the integration of the BPS 
eqs.(\ref{newbps1})-(\ref{newbps4}) for  the 
type {\bf A} backgrounds (\ref{metricA}).

We start by noting that the difference $H-G$ can be integrated immediately,
\be
G= H + \frac {N_f -2N_c}{2} \rho - C 
\label{GofH}
\ee
where $C$ is a constant of integration.
The function $Y$ can also be written in terms of $H$,
\be
Y= \frac12 \partial_\rho H +\frac14 (N_f - N_c)
\label{YfromH}
\ee
This allows us to write the system as a single 
second order differential equation for $H$,
\be
\frac{\partial^2 H}{\partial\rho^2}=
\left(\frac12 \partial_\rho H + \frac14 (N_f - N_c)\right)
\left[-2\frac{\partial _\rho H + N_f - N_c}{H}-
\frac{N_f + 2 \partial _\rho H}{H + \frac{N_f -2N_c}{2} \rho - C} + 8
\right]
\label{eqforH}
\ee
and the dilaton that can be obtained as,
\beq
\phi(\rho)=\phi_0 +\int \Big[\frac{N_f-N_c}{4H}+\frac{N_c}{4G}   
\Big]d\rho
\label{dilofH}
\eeq
There is of course, one simple solution for $N_f = 2N_c$ (the case $N_f = 
2 N_c$ is the "conformal solution" studied 
in \cite{Casero:2006pt}) that reads
$H=N_c/\xi$, $G= N_c/ (4-\xi)$, $Y=N_c/4$, but even in the 
particular case $N_f=2N_c$ there are  more solutions that we will not 
discuss here \cite{Caceres:2007mu}. Unfortunately we did not succeed in 
finding 
other exact 
solutions to 
(\ref{eqforH}). Nonetheless we derive below the 
asymptotic expansions, near the UV 
and the IR, of the solutions and then show that a 
smooth interpolation between them exists. 
\subsection{UV expansion ($\rho\to\infty$)}\label{sec: UV}
As explained above, we are interested in solutions to 
(\ref{newbps1})-(\ref{newbps4}) 
with asymptotically linear dilaton.
There are two distinct possibilities, depending on 
whether $N_f$ is bigger or smaller than 
$2N_c$. 
\subsubsection{$N_f < 2N_c$}
There is a one-parameter family (depending on the constant $C$ 
that enters (\ref{GofH})) of
UV (large $\rho$) asymptotic solutions. For the different functions we have:
\bear
H& =& \left(N_c - \frac{N_f}{2}\right)\rho + \left(C + \frac{N_c}{4}\right) + \frac{N_c N_f}{16(2N_c - N_f)}\rho^{-1}+ \dots\rc
G &=& \frac{N_c}{4} + \frac{N_c N_f}{16(2N_c - N_f)} \rho^{-1}  + \frac{N_c N_f (N_f -2N_c -4C)}{32 (2 N_c - N_f)^2}\rho^{-2}+\dots\rc
Y &=&  \frac{N_c}{4}  - \frac{N_c N_f}{32(2N_c - N_f)} \rho^{-2} -  \frac{N_c N_f (N_f -2N_c -4C)}{32 (2 N_c - N_f)^2}\rho^{-3}+\dots\rc
\partial_\rho \phi &=& 1- 4 \rho^{-1} + \frac{8C + 2N_c + 2N_f}{16 (2N_c -N_f)}\rho^{-2} + \dots
\label{UVsmallNf}
\eear
This coincides with the asymptotic expansion 
found in equation (4.19) of \cite{Casero:2006pt} for the solutions of type {\bf 
N}, except 
for the fact
that here there is an extra free parameter $C$, which in  
those solutions  seems to be fixed to $C= (N_f -2N_c)/4$.
\subsubsection{$N_f > 2 N_c$}
Here again a one-parameter family of solutions depending on $C$ is 
found,
\bear
H&=&\frac{N_f -N_c}{4} + \frac{N_f (N_f - N_c)}{16(N_f -2N_c)}\rho^{-1} +  \frac{N_f (N_f - N_c)(4C + 2N_c - N_f)}{32 (N_f - 2N_c)^2}\rho^{-2}+\dots \rc
G&=& \frac{N_f-2N_c}{2}\rho + \frac{N_f - N_c - 4C}{4} +  \frac{N_f (N_f - N_c)}{16(N_f -2N_c)}\rho^{-1} +\dots \rc
Y&=& \frac{N_f - N_c}{4} -  \frac{N_f (N_f - N_c)}{32(N_f -2N_c)}\rho^{-2} -  \frac{N_f (N_f - N_c)(4C + 2N_c - N_f)}{32 (N_f - 2N_c)^2}\rho^{-3}+\dots \rc
\partial_\rho \phi &=& 1- 4 \rho^{-1} + \frac{-8C - 2N_c + 3N_f}{16 (N_f-2N_c)}\rho^{-2} + \dots
\label{UVbigNf}
\eear
which coincides with the expansion for the type {\bf N } backgrounds 
found in eq.(4.20) of \cite{Casero:2006pt} if 
we  set $C= (N_f -2N_c)/4$.
\subsection{IR behaviour}
\label{sec: IR}
Let us now present the solutions to the BPS eqs. at small
$\rho$, as they appear from 
solving (\ref{eqforH}) together with 
(\ref{GofH}), (\ref{YfromH}) and (\ref{dilofH}). 
We remind the reader that, as discussed in the beginning of section 
\ref{sectionsolutions}, we require $N_f \geq N_c$
so it is guaranteed that the quantity $g_{tt}\sim 
e^{\frac{\phi}{2}}$ does not diverge in the IR. 
The space is singular at the origin, but we will call 
it a good singularity and 
proceed assuming that correct physics predictions for the IR of our  
theory 
of interest ({\ref{ourtheory}}) can be obtained by doing computations with 
the type {\bf A} backgrounds.

There is  quite a rich structure since we actually 
found three possible IR expansions that can be smoothly connected with the UV 
expansions of section \ref{sec: UV}.  The three possible IR 
behaviors will be labelled
as type I, type II and type III. Let us study them in turn.

\subsubsection{Type I expansion}\label{typeiexpansionss}

An obvious solution of (\ref{eqforH}) is $H= \frac12 (N_f - N_c) (c_1 - \rho)$, 
where $c_1$ is an integration constant. But eq.
(\ref{YfromH}) would yield $Y=0$. Thus, this simple solution cannot be 
physical. 
Nevertheless, there are solutions that asymptotically approach this behaviour in the
IR (as $\rho \to -\infty$). The deviation from the simple linear solution is given by
exponentially suppressed terms (as $\rho \to -\infty$)  in the 
following way,
\be\label{HIR}
H(\rho)= \frac{N_f-N_c}{2}(c_1 - \rho) +\sum_{k\geq 1} {\mathcal P}_k(\rho)e^{4k\rho}
\qquad  (\rho \to -\infty)\,,
\ee
The ${\mathcal P}_k(\rho)$ are order $k+1$ polynomials in $\rho$.
The first one is
\be
{\mathcal P}_1 = c_3 \left[ \rho^2 -\rho (c_1 + c_2 + \frac12)
+ c_1  c_2 + \frac14 (c_1 + c_2) + \frac18 \right]
\ee
and the higher ones can be obtained iteratively.
Notice that one of the integration constants can be reabsorbed by shifting the origin of $\rho$
and that there is a relation among the different constants 
$C= \frac{c_1}{2} (N_f -N_c) - \frac{c_2}{2} N_c$.
The leading IR behaviour of the rest of the functions can be readily obtained
\bear
G(\rho)&=& \frac{N_c}{2}(c_2 - \rho) + {\cal O}(e^{4\rho}) 
 \rc
Y(\rho)&=& 4 c_3 e^{4\rho} (c_1 - \rho)(c_2 - \rho)
+ {\cal O}(e^{8\rho}) \qquad (\rho \to - \infty) \rc
e^\phi &=& \frac{e^{\phi_0}}{ \sqrt{(c_1 - \rho)(c_2 - \rho)}}
+ {\cal O} (e^{4\rho})
\eear
This IR asymptotics can be numerically connected to the UV behaviours 
described in section (\ref{sec: UV}).
We depict  some sample plots in section \ref{sec: numericplots}.

\subsubsection{Type II expansions}\label{typeiiexpansionss}

Differently from the type I expansion we presented in the previous subsection, equation (\ref{eqforH}) also admits solutions where $H$, $G$, or both vanish at a finite value of $\rho$, 
which we take to be $\rho=0$ without loss of generality.
 The way $H$ and $G$ reach the origin depends essentially on the value of the constant $C$ in (\ref{eqforH}). When $C$ is strictly negative $H$ goes to zero like  $\rho^\frac{1}{2}$ whereas $G$ goes to a constant. When $C$ is strictly positive, $H$ and $G$ exchange roles: $H$ goes to a constant while the leading behavior of $G$ is $\rho^\frac{1}{2}$. We call both these kinds of solutions type II. When $C=0$, instead, $H$ and $G$ vanish at the same time, and their leading term goes like $\rho^\frac{1}{3}$; we call this solution type III and postpone its presentation until the next subsection.

\begin{figure}[htbp] %  figure placement: here, top, bottom, or page
  \centering
  \includegraphics[width=0.45\textwidth]{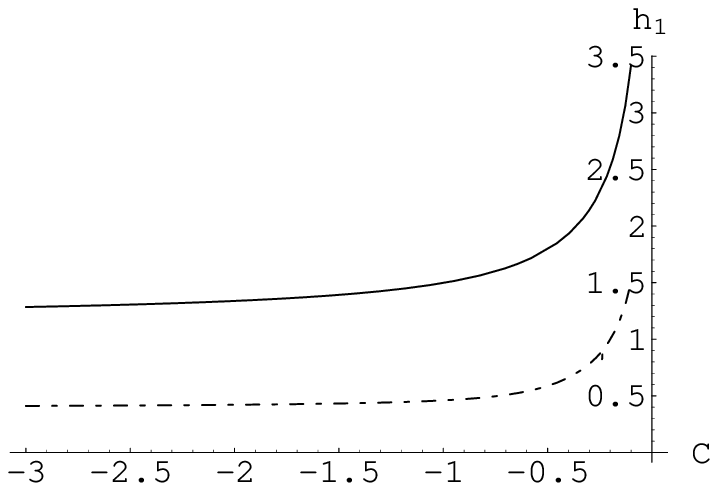}\hfill
  \includegraphics[width=0.45\textwidth]{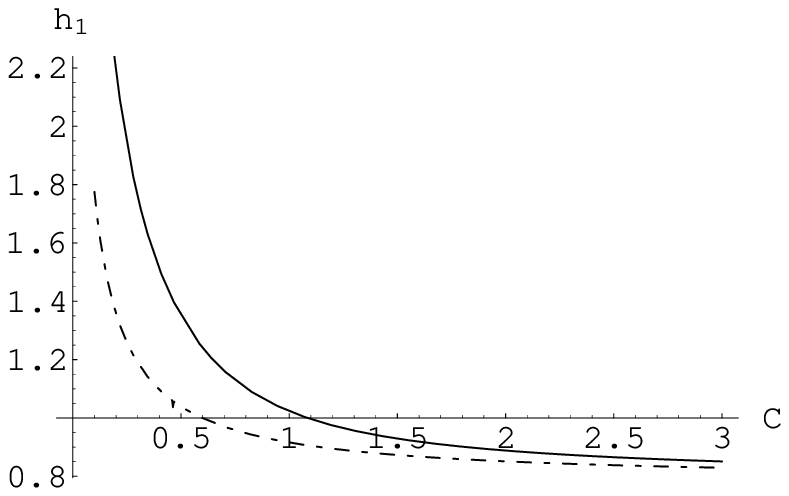}
  \caption{We represent here the relationship between $C$ and $h_1$ that ensures that the evolution of a given type II expansion goes asymptotically to the behaviour we presented in section \ref{sec: UV}. On the left we are considering negative values of $C$, corresponding to expansion (\ref{type II neg}) while on the right positive values are considered, which correspond to (\ref{type II pos}). Dashed lines correspond to $\frac{N_f}{N_c}=1.5$ while continuous lines correspond to $\frac{N_f}{N_c}=2.5$.}
  \label{fig: C-h1 type II}
\end{figure}

More explicitly, for $C<0$ the following expansions solve equations (\ref{GofH}) - (\ref{dilofH})
\begin{align}
&H=h_1\rho^\frac{1}{2} +\left(\frac{h_1^2}{3C}+N_c-N_f\right) \rho + h_1 \frac{72 C^2 + 20 h_1^2+ 3C (10N_c-7N_f)}{72C^2}\rho^\frac{3}{2}+\ldots\nonumber\\
&G=-C +h_1\rho^\frac{1}{2}+\left(\frac{h_1^2}{3C}-\frac{N_f}{2}\right) \rho+ h_1 \frac{72 C^2 + 20 h_1^2+ 3C (10N_c-7N_f)}{72C^2}\rho^\frac{3}{2}+\ldots\nonumber\\
&Y=\frac{h_1}{4\rho^\frac{1}{2}} +\frac{1}{12}\left(\frac{2h_1^2}{C}+3 N_c -3 N_f\right) + \frac{3}{4}h_1  \frac{72 C^2 + 20 h_1^2+3 C (10N_c-7N_f)}{72C^2}\rho^\frac{1}{2}+\ldots\nonumber\\
&\phi=\phi_0 + \frac{N_f-N_c}{2h_1}\rho^\frac{1}{2} + \frac{3C(N_f-N_c)^2-h_1^2(2N_c+N_f)}{12C h_1^2}\rho+\ldots \label{type II neg}
\end{align}
whereas for $C>0$, we have the following power series
\begin{align}
&H=C +h_1\rho^\frac{1}{2}-\left(\frac{h_1^2}{3C}+\frac{N_f}{2}\right) \rho+ h_1 \frac{72 C^2 + 20 h_1^2+ 3C (10N_c-3N_f)}{72C^2}\rho^\frac{3}{2}+\ldots\nonumber\\
&G=h_1\rho^\frac{1}{2}-\left(\frac{h_1^2}{3C}+N_c\right) \rho+ h_1 \frac{72 C^2 + 20 h_1^2+ 3C (10N_c-3N_f)}{72C^2}\rho^\frac{3}{2}+\ldots\nonumber\\
&Y=\frac{h_1}{4\rho^\frac{1}{2}}-\frac{1}{2}\left(\frac{h_1^2}{3C}+\frac{N_c}{2}\right)+h_1\frac{72C^2+20h_1^2+3C(10N_c-3N_f)}{96 C^2}\rho^\frac{1}{2}+\ldots\nonumber\\
&\phi=\phi_0+\frac{N_c}{2h_1}\rho^\frac{1}{2}+\frac{3C+h_1^2(3N_f-2N_c)}{12 C h_1^2}\rho+\ldots \label{type II pos}
\end{align}
The constants $h_1$, $C$ and $\phi_0$ are arbitrary. 
In particular $\phi_0$ can be fixed to any value. 
For what concerns $C$ and $h_1$, instead, what happens is more 
interesting. While any choice of them gives a solution to 
equations (\ref{GofH}) - (\ref{dilofH}), requiring that a 
particular solution behaves in the UV as  the functions in 
section \ref{sec: UV} imposes an additional condition that 
enforces a relation between the useful values of $C$ and $h_1$. 
Figure \ref{fig: C-h1 type II} is the plot of such a 
relation for two fixed values of $\frac{N_f}{N_c}$. 
All other possible choices still give a solution to 
(\ref{GofH}) - (\ref{dilofH}), but the functions either
 reach a singularity at some finite value of $\rho$ or 
 reach the UV with a different asymptotical behaviour, in
 which we are not interested here (analogous to the
 one studied in section 8 of \cite{Casero:2006pt}).

\subsubsection{Type III expansion}\label{typeiiiexpansionss} 

When the constant $C$ in (\ref{GofH}) vanishes, type II expansions are 
not allowed anymore, but a new branch of solutions opens up, which we will call type III. In this case both $H$ and $G$ vanish at the origin of space. The expansion reads as follows
\begin{align}
&H=h_1\rho^\frac{1}{3} +\frac{1}{10}(5N_c-7N_f)\rho+\frac{2}{3}h_1\rho^\frac{4}{3}+\ldots\nonumber\\
&G=h_1\rho^\frac{1}{3} -\frac{1}{10}(5N_c+2N_f)\rho+\frac{2}{3}h_1\rho^\frac{4}{3}+\ldots\nonumber\\
&Y=\frac{h_1}{6}\rho^{-\frac{2}{3}}-\frac{N_f}{10}+\frac{4}{9}h_1\rho^\frac{1}{3}+\ldots\nonumber\\
&\phi=\phi_0+\frac{3N_f}{8h_1}\rho^\frac{2}{3}+3\frac{10N_c^2-10N_cN_f+7N_f^2}{160h_1^2}\rho^\frac{4}{3}+\ldots
\end{align}
Again the constants $h_1$ and $\phi_0$ are arbitrary, but to obtain a solution that asymptotes to the behaviour of section \ref{sec: UV}, we have to choose a particular 
value of $h_1$ for any fixed value of $\frac{N_f}{N_c}$.
\subsection{Numerical solutions}
\label{sec: numericplots}

\begin{figure}[tb] %  figure placement: here, top, bottom, or page
   \centering
   \includegraphics[width=0.45\textwidth]{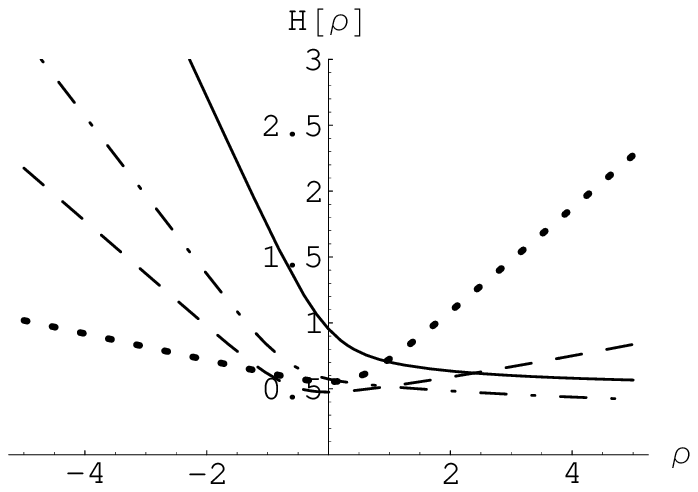}\hfill  \includegraphics[width=0.45\textwidth]{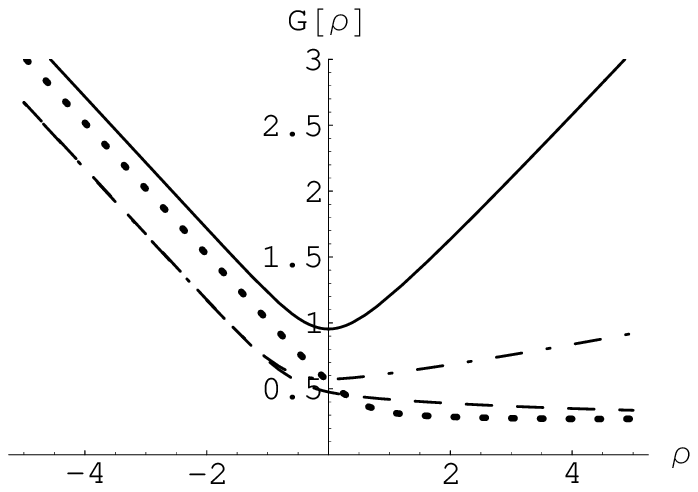}\\[3pt]
    \includegraphics[width=0.45\textwidth]{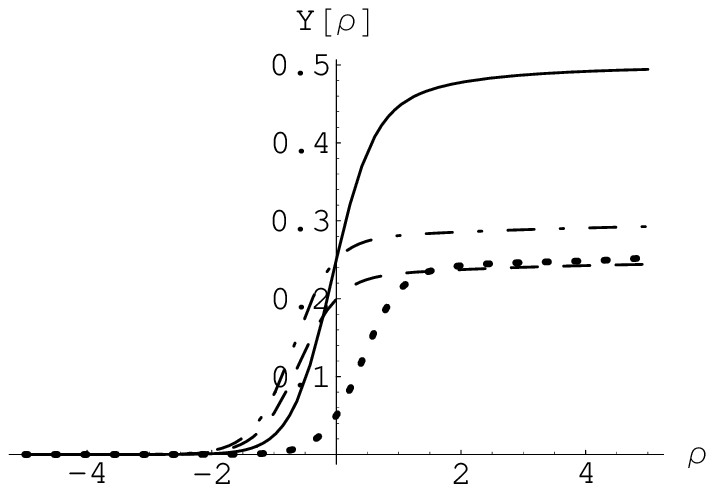}\hfill  \includegraphics[width=0.45\textwidth]{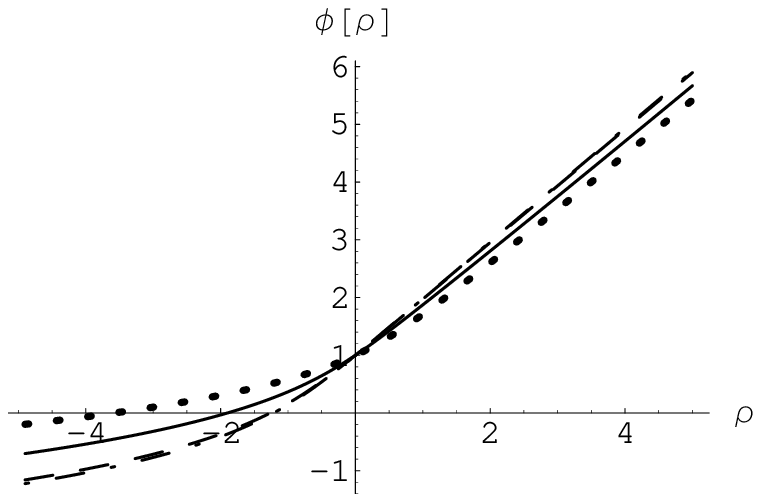}
   \caption{Type I solutions obtained numerically. We have fixed $C=0$, $N_c=1$ and defined $\rho=0$ as the
point where $H$ ($G$) reaches a minimum for $N_f <2N_c$ ($N_f > 2N_c$).
We plot the solutions for
 different values of $\frac{N_f}{N_c}$: dotted, dashed, dot-dashed and continuous lines correspond respectively to $N_f=1.2$, $1.8$, $2.2$, $3$ $N_c$.}
   \label{fig: type I}
\end{figure}
The expansions above are essential in the analysis of solutions to the system (\ref{GofH})-(\ref{dilofH}), but their knowledge is not enough to guarantee the existence of an overall solution that connects the given UV and IR behavior. We have checked that such connecting solution
indeed exists for the three kinds of IR behaviours. In this section, we plot some sample numerical computations.

Figure \ref{fig: type I} gives some examples of the numerical integration for the type I IR behaviour.

\begin{figure}[htbp] %  figure placement: here, top, bottom, or page
   \centering
   \includegraphics[width=0.45\textwidth]{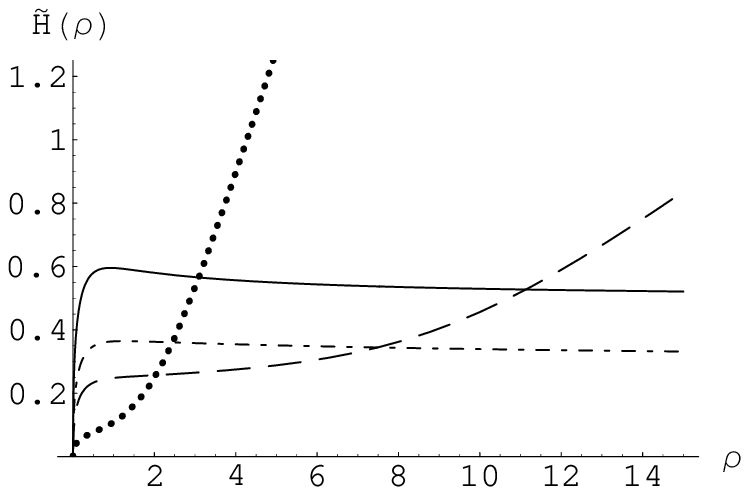}\hfill  \includegraphics[width=0.45\textwidth]{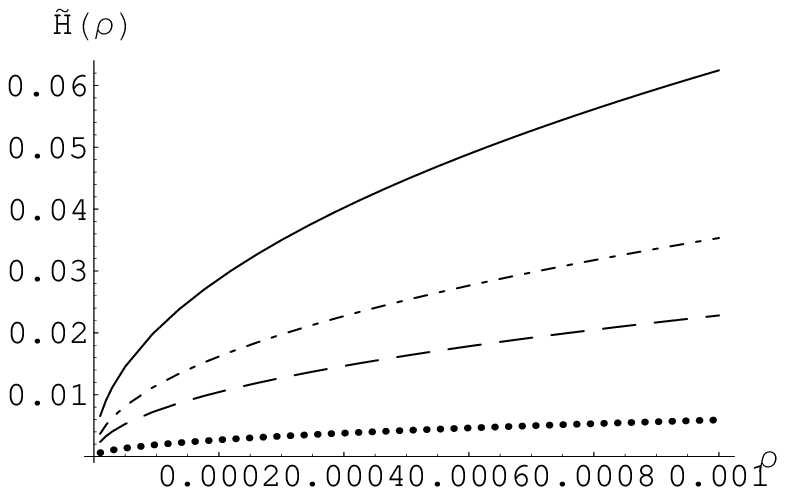}\\[3pt]
    \includegraphics[width=0.45\textwidth]{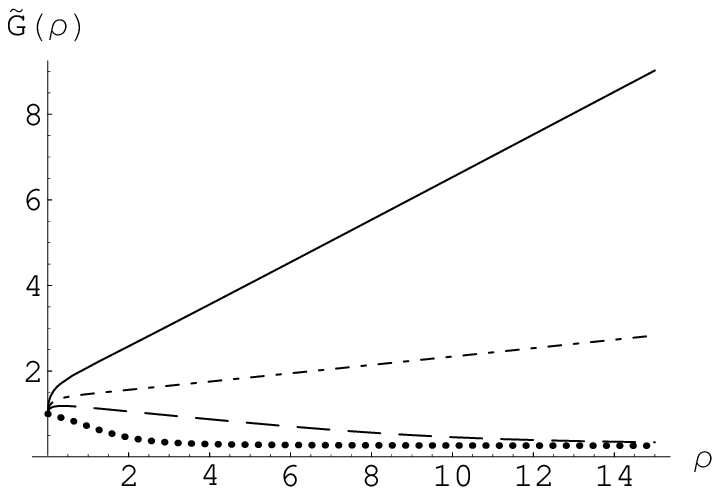}\hfill  \includegraphics[width=0.45\textwidth]{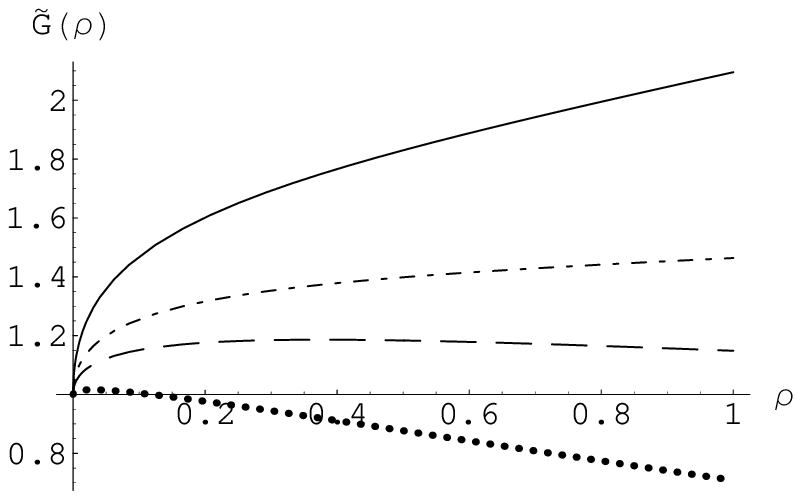}\\[3pt]
        \includegraphics[width=0.45\textwidth]{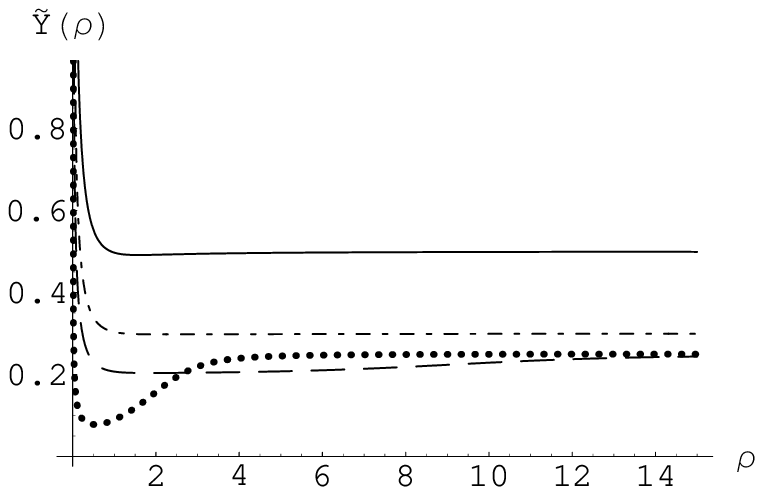}\hfill  \includegraphics[width=0.45\textwidth]{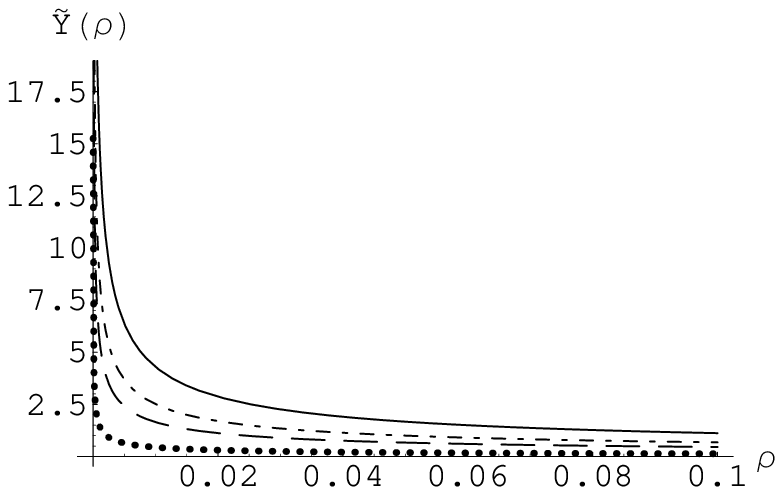}\\[3pt]
    \includegraphics[width=0.45\textwidth]{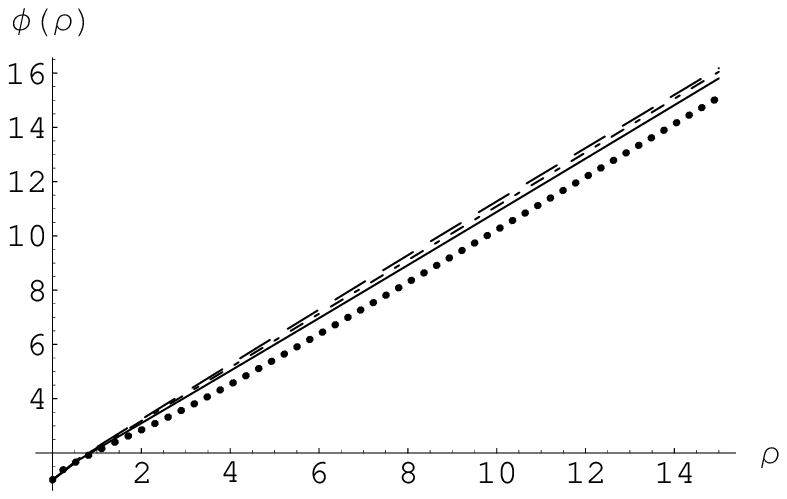}\hfill  \includegraphics[width=0.45\textwidth]{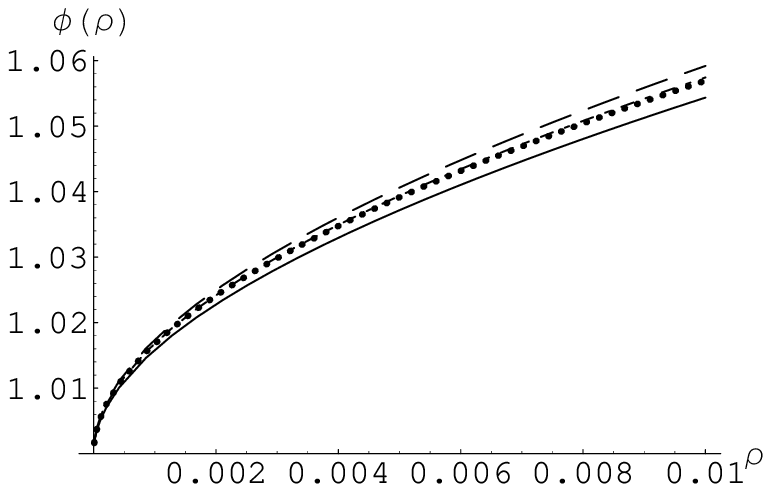}
   \caption{Type II solutions derived numerically. In these 
   plots we present the results for $C=-1$ and different values 
   of $\frac{N_f}{N_c}$: dotted, dashed, dot-dashed and continuous 
   lines correspond respectively to $N_f=1.2$, $1.8$, $2.2$, $3$ $N_c$. 
   The graphs on the left-hand side represent the solutions for the 
   background on a large range of $\rho$, while the plots on the 
   right-hand column focus on the behaviour of the functions close 
   to the origin. Tilded function are rescaled by $N_c$ as
   $\tilde H = \frac{H}{N_c}$, etc.}
   \label{fig: type II C<0}
\end{figure}

\begin{figure}[htbp] %  figure placement: here, top, bottom, or page
   \centering
   \includegraphics[width=0.45\textwidth]{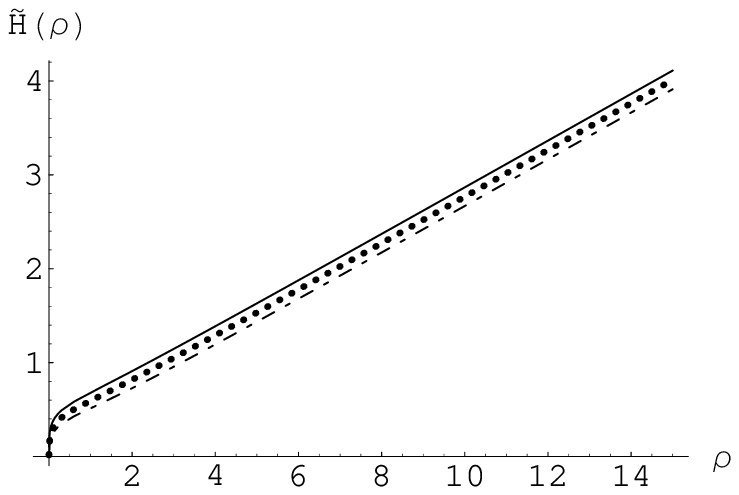}\hfill  \includegraphics[width=0.45\textwidth]{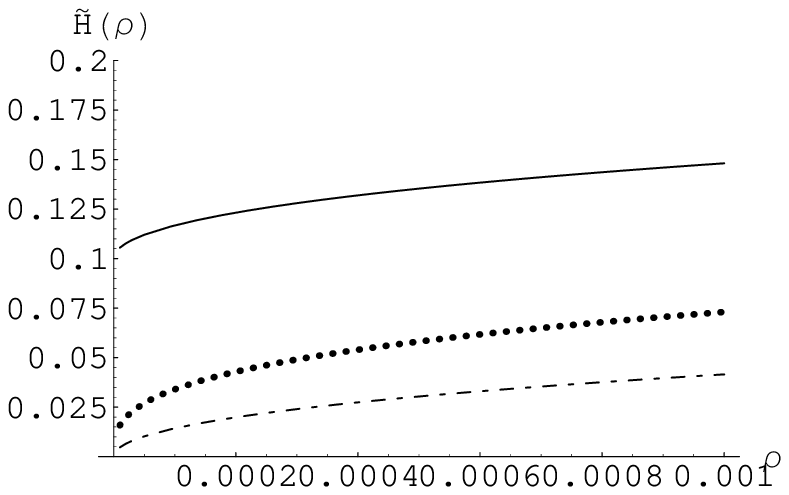}\\[3pt]
    \includegraphics[width=0.45\textwidth]{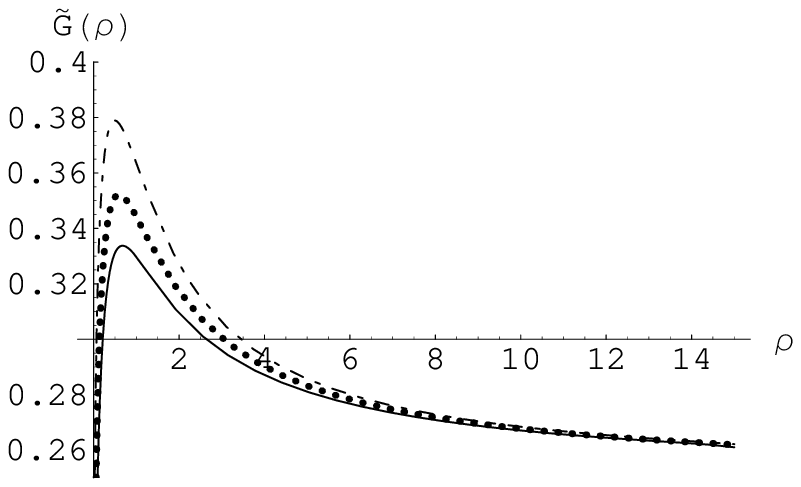}\hfill  \includegraphics[width=0.45\textwidth]{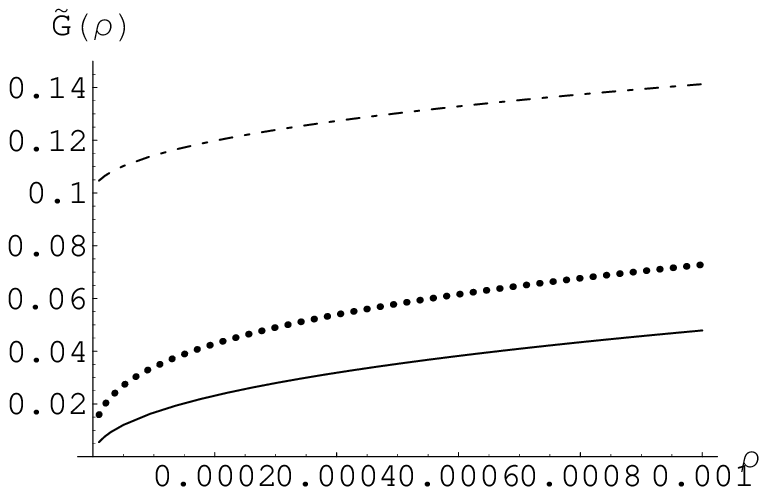}\\[3pt]
        \includegraphics[width=0.45\textwidth]{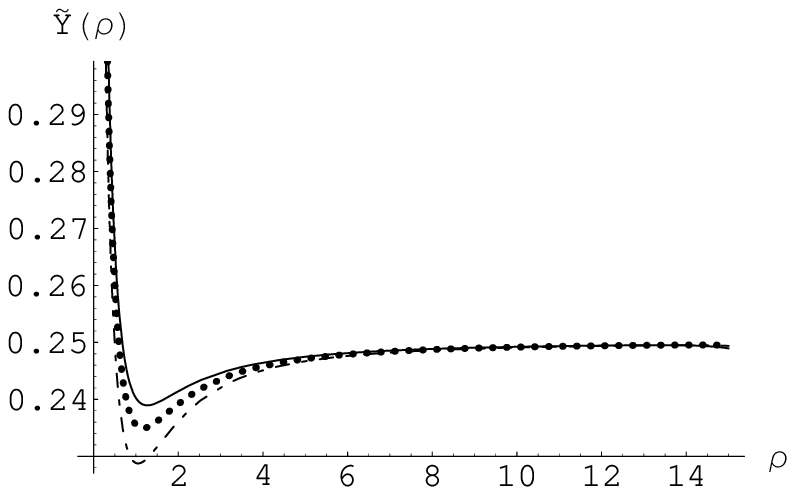}\hfill  \includegraphics[width=0.45\textwidth]{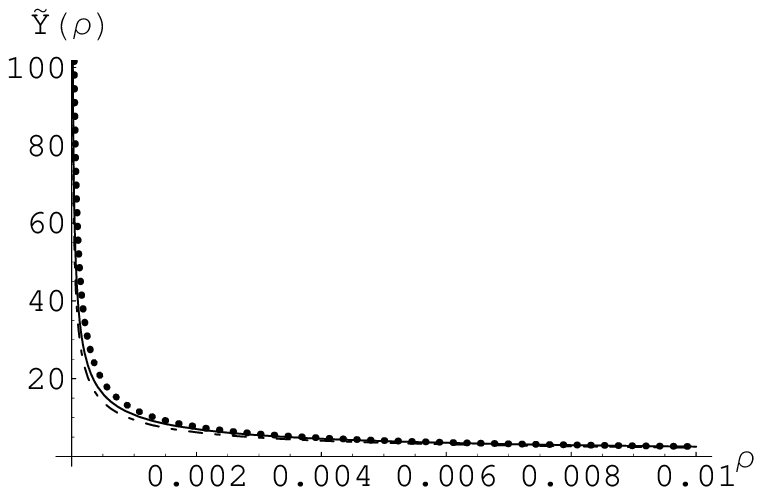}\\[3pt]
    \includegraphics[width=0.45\textwidth]{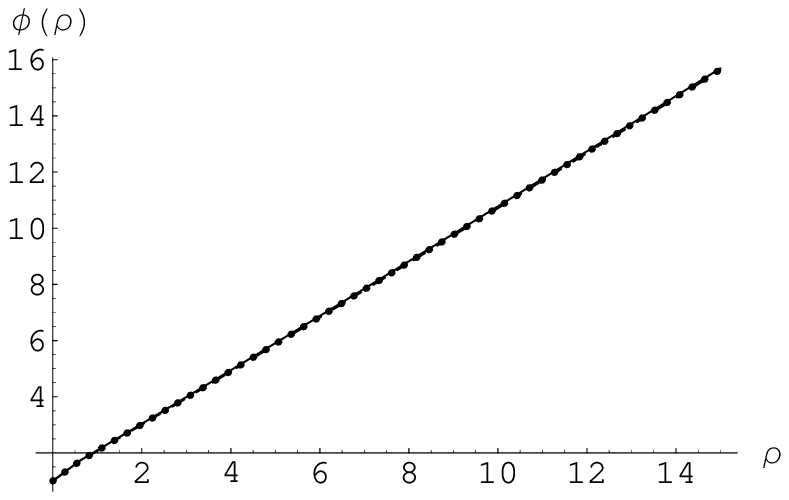}\hfill  \includegraphics[width=0.45\textwidth]{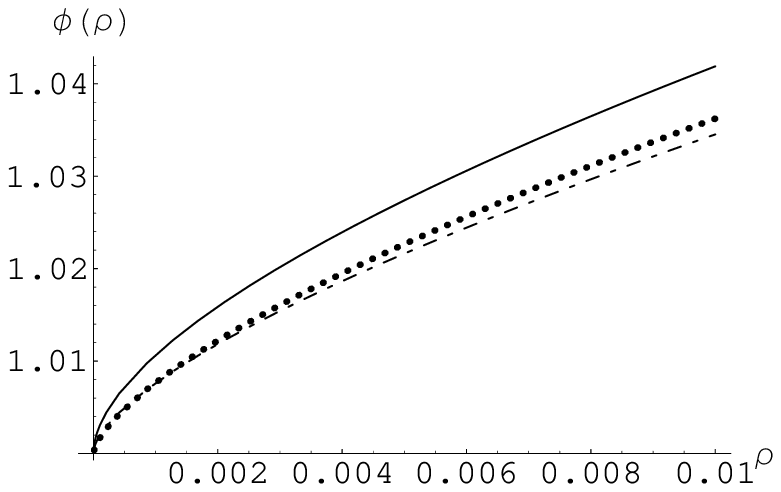}
   \caption{Type II and III solutions derived numerically. 
   In these plots we present the behaviour of type III solutions, 
   and compare it with small-$C$ type II solutions. Clearly the 
   transition through $C=0$ is almost continuous. Here we have 
   fixed $N_f=1.5 N_c$. Dotted lines correspond to the type III 
   solution, while dot-dashed and continuous lines correspond to 
   type II backgrounds with respectively $C=-0.1$ and $C=0.1$.}
   \label{fig: type III}
\end{figure}

As we have already anticipated, requiring that the type II and III expansions are connected with the UV asymptotics of section \ref{sec: UV} imposes an additional condition, fixing the value of the integration constant $h_1$ in terms of $C$ and $\frac{N_f}{N_c}$. 
This additional condition produces the plots in Figure \ref{fig: C-h1 type 
II}.
Given that we choose the suitable value of $h_1$,  
it is then possible to numerically  integrate equations 
(\ref{GofH})-(\ref{dilofH}), and find a complete solution for the 
type {\bf A} backgrounds. For type II and type 
III solutions, we present examples of these solutions in 
Figures \ref{fig: type II C<0} and \ref{fig: type III}, for different 
values of $C$ and $\frac{N_f}{N_c}$.

\subsection{UV expansions revisited}
\label{sec: revisit}

The reader may wonder how it is possible that different IR behaviours asymptote to the
same UV when it seems that the UV expansions (\ref{UVsmallNf}), (\ref{UVbigNf}) have
no free parameters apart from $C$. Indeed, those expansions are not complete and
there is an extra free parameter in the UV if one
considers the possibility of exponentially suppressed terms: we 
can 
write,
\be
H= Q_0(\rho) + e^{-4\rho} \rho^{-\alpha} Q_1(\rho) + {\cal O} (e^{-8\rho})
\ee
where $Q_0$ are the expansions (\ref{UVsmallNf}), (\ref{UVbigNf})
with:
\bear
\alpha = \frac{2N_c - 3N_f}{3(2N_c - N_f)}\qquad (N_f <2N_c) \rc
\alpha = \frac{2N_c + N_f}{3(2N_c - N_f)}\qquad (N_f > 2N_c)
\eear
The $Q_i(\rho)$ are polynomials such that
$Q_i(\rho) = Q_{i,0} + Q_{i,1} \rho^{-1} +  Q_{i,2} \rho^{-2}+\dots$.
$Q_{1,0}$ is a free parameter and the rest are fixed in terms of it.
We have, thus (once $C,N_c,N_f$ are fixed), a one parameter family of
UV solutions, labelled by $Q_{1,0}$. 
Different choices $Q_{1,0}$ should modify the IR behaviour and, in particular, make the system
fall into one of the qualitatively different IR possibilities. However, checking this numerically
requires great numerical precision and we could not bear 
out this expectation explicitly.

It would be interesting to know the physical interpretation of this parameter
$Q_{1,0}$. Since it comes multiplying a term with $e^{-4\rho}$, it must be related to
the vacuum expectation value of an operator of UV dimension 6, here we 
anticipate the
UV radius-energy relation $\log[\mu/\Lambda]= \frac{2}{3}\rho$ that we will use in equation
(\ref{re}). The simplest guess is an operator of the type $(Q \tilde Q)^4$ 
(since in the UV $\gamma_Q \to - \frac12$, at least when $N_f \leq 2N_c$).
A single vev of this type would spontaneously break the R-symmetry, apparently in contradiction
with the described solution. However, since there are infinitely many smeared flavored branes in the
construction, it is natural to think that, if in fact such vevs are turned on, they may also be smeared
effectively restoring the $U(1)_R$. This is similar to the discussion in  \cite{Paredes:2006wb} where,
considering smeared masses for hypermultiplets, the R-symmetry that would be broken for a single mass
is effectively restored.

\subsection{A comment on integration constants}

The different expansions are written in terms of some
integration constants which are related among them.
In order to avoid confusion, we clarify here the
number of parameters on which the different families of solutions
depend.

We have four first order equations (\ref{newbps1})-(\ref{newbps4}).
Let us fix $N_f / N_c$ (notice that, mantaining 
this ratio fixed, $N_c$ can be reabsorbed
by simultaneously rescaling $H,G,Y$).
Then a solution of the system depends on four integration constants. 
One is an additive
constant for $\phi$ that does not affect the integration of the other
functions. Since the BPS eqs do not depend explicitly on $\rho$, a second
integration constant is just a shift of $\rho$, which we can fix without
loss of generality. In type I, we can define $\rho=0$ as the point
where $H'$ (or $G'$) vanishes, as in figure \ref{fig: type I}.
For type II and III, we have defined $\rho=0$ as the point where
$H$ (and/or $G$) vanish. The third integration constant is fixed
by requiring that the system goes to the linear dilaton UV
(so there are not growing exponentials in $H$). From the IR point of
view, this is, for instance, what fixes $h_1$ in terms of $C$ in the
type II IR expansions, see the discussion in section \ref{typeiiexpansionss}.

Thus, for fixed $N_f/N_c$ and fixing also the additive constant of $\phi$,
we are left with the fourth integration constant,
which gives
a one-parameter family of
type I and type II solutions. For type III, where we further
require $C=0$, there is a single solution
(a limiting point of the type II family).

After these discussions on the solutions, we will focus on obtaining 
physics predictions from them. To this we turn now.

\section{Physics predictions of the new solutions}\label{predictions}
\setcounter{equation}{0}
In this section, we study some non-perturbative aspects that can be 
learned from the solutions presented in section \ref{sectionsolutions}. We 
will 
discuss expressions for the gauge coupling and
anomalous dimensions of the quark superfields,  give predictions for 
their values and compute the 
beta function in the UV of the theory (matching 
results 
known from field theory). The predicted anomalous dimension for the quark 
superfield will be in agreement with the results of eq.(\ref{gammauv}) and 
those spelled out  in section \ref{anomalymatching}. We will present 
similar results for the beta function of the quartic coupling (for which 
we will provide a geometrical definition). We will study the Wilson and 't 
Hooft loops and address in detail the issue of (continuous and discrete) 
anomaly matching.

Let us remind the 
reader that we take units so that 
$\alpha'=g_s=1$. With this choice, the tension of a D5 brane is 
$T_5=\frac{1}{(2\pi)^5}$. 
\subsection{Theta-angle and gauge coupling}
\label{sec:thetagym}
We will give an
expression for the theta-angle and gauge coupling that can be simply 
derived by 
considering the action for a D5 brane on the submanifold
$(t,\vec{x}, \theta=\tilde\theta, \varphi=2\pi-\tilde\varphi)$, at 
constant radial coordinate. This is a well known definition, originally 
proposed in \cite{Di Vecchia:2002ks}. In \cite{Casero:2006pt} we defined 
the gauge coupling 
using an instanton, whose action is computed using  a D1 brane wrapping 
the two-cycle $(\theta=\tilde\theta, \varphi=2\pi-\tilde\varphi)$. This 
two-manifold is special, being the only shrinking closed manifold in the 
geometry. As expected, the results using D5 or D1 probes do coincide.

We consider the dynamics of a D5 with gauge fields 
$A_\mu $ excited on the $R^{1,3}$ directions, obtaining (in string frame),
\beq
S_{D5}= -T_{5}\int d^6 x e^{-\phi} \sqrt{-\det(g+ 2\pi F)} + 
\frac12 (2\pi)^2 T_5 
\int 
C_2\wedge 
F_2\wedge F_2
\label{bigauge}
\eeq
The induced metric and two-form 
on the probe D5 brane  read, in 
string frame,
\bea
ds^2_{ind}&=& e^\phi \Big[ dx_{1,3}^2 + (H+G)(d\theta^2 +\sin^2\theta 
d\varphi^2)  \Big]\,,\rc
(C_2)_{ind} &=& \frac{2N_c - N_f}{4}(\psi - \psi_0) \sin \theta d\theta \wedge
d\varphi
\eea
Following the procedure of \cite{Di Vecchia:2002ks},
we can expand (\ref{bigauge}) up to second order in the gauge fields (this 
is the QFT 
part of the $\alpha'$ expansion), perform the integrals over the 
two-cycle and compare to the usual Yang-Mills action in order
to obtain the couplings.

For the theta-angle, we get:
\beq
\Theta=\frac{(2N_c-N_f)}{2}(\psi-\psi_0).
\label{Theta-angle}\eeq
Since changes in $\Theta$ are associated with traslations $\psi\to 
\psi+2\epsilon$ then only some values of $\epsilon= 
\frac{2k\pi}{2N_c-N_f}$ are allowed and they correspond to changes in 
$\Theta\to \Theta+2k\pi$, which nicely matches what we wrote about 
anomalies between eq.(\ref{c2})-(\ref{match}). 

\begin{figure}[htb] %  figure placement: here, top, bottom, or page
   \centering
   \includegraphics[width=0.45\textwidth]{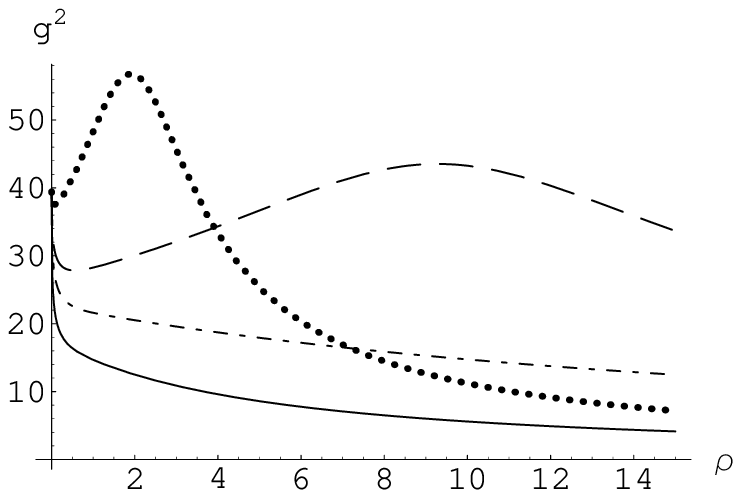}\hfill \includegraphics[width=0.45\textwidth]{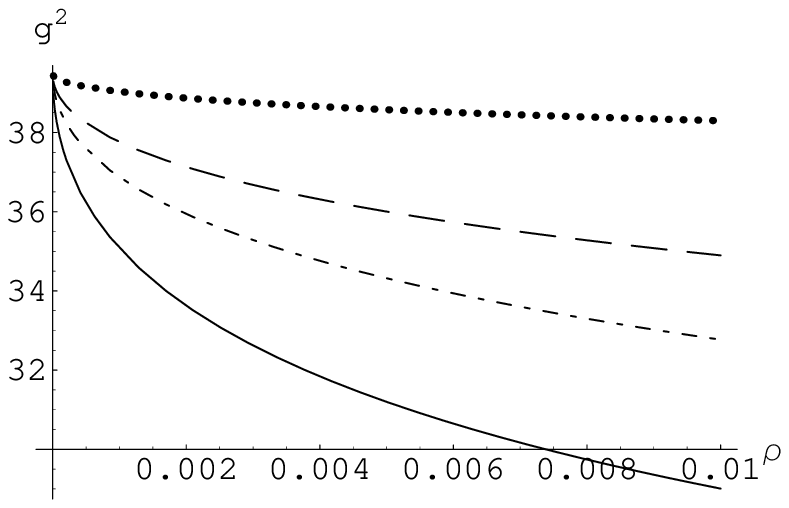}\\[3pt]
   \includegraphics[width=0.45\textwidth]{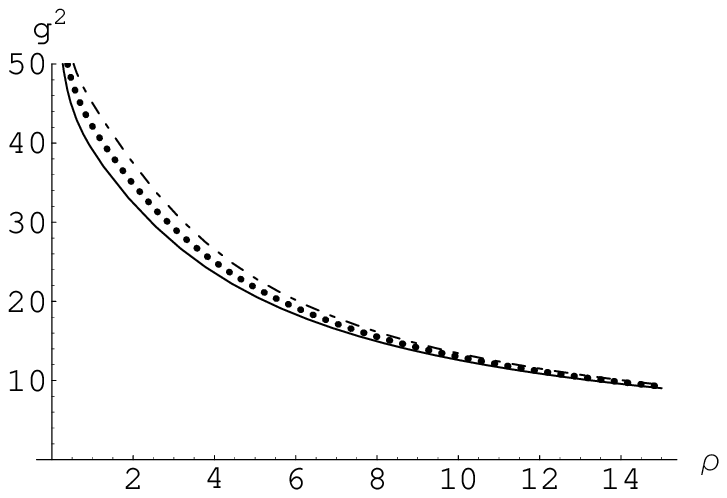}\hfill  \includegraphics[width=0.45\textwidth]{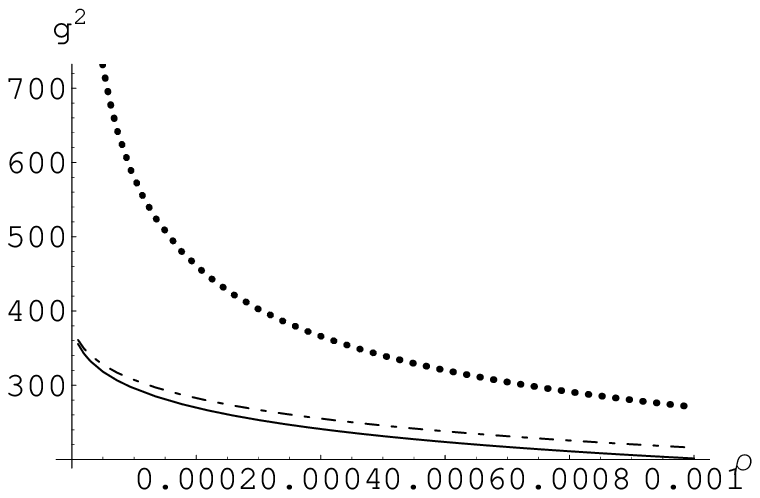}
   \caption{These plots represent the behaviour of the gauge 
   coupling as defined in equation (\ref{esta}) in 
   type II (top) and type III  (bottom) backgrounds. 
   For the top plots we have taken $C=-1$ and different 
   values of $\frac{N_f}{N_c}$. The correspondence between line 
   styles and $\frac{N_f}{N_c}$ is the same as in figure 
   \ref{fig: type II C<0}. For the bottom plots, 
   we have taken $N_f=1.5 N_c$ and $C=-0.1$ and $0.1$ (type II), 
   and $C=0$ (type III). For the correspondence between values 
   of $C$ and line styles we refer the reader to the legend of 
   figure \ref{fig: type III}. The plots on the right are a zoom
   around $\rho=0$ of the plots on the left.}
   \label{fig: gauge}
\end{figure}

The gauge coupling in our SQCD theory takes 
the form,
\beq
\frac{8\pi^2}{g^2} =2(H+G) .
\label{esta}
\eeq
For type I solutions, the series expansions of section 
\ref{typeiexpansionss} show 
that the gauge coupling vanishes in the IR. The expansions of section 
\ref{typeiiexpansionss} for type II solutions show a version 
of `soft confinement' (with bounded gauge coupling in the IR), while for 
type III more 
conventional confinement is obtained (with divergent gauge coupling in the IR).
Figure \ref{fig: gauge} shows the behavior of the coupling as a function 
of the radial coordinate. 
\subsubsection{Beta Function and Anomalous Dimension}\label{betaanom}
We can compute the beta function of the gauge coupling as defined in 
(\ref{esta}). As we stated below eq.(\ref{betag}), we will be interested 
in the wilsonian beta 
function, 
that takes the expression,
\beq
\beta_{\frac{8\pi^2}{g^2}} = \frac{\partial\Big( 
\frac{8\pi^2}{g^2}\Big)}{\partial 
\log(\mu/\Lambda)}= 3 N_c - N_f(1-\gamma_Q).
\label{betain}\eeq 
Now, we need a relation between the radial coordinate $\rho$ and the scale 
$\mu$ in order to perform the derivative in (\ref{betain}). In  
section 5.5 of \cite{Casero:2006pt} we proposed
that the radius-energy relation, in the UV of the gauge theory, that is 
for large values of the radial coordinate, is:
\beq
\log[\mu/\Lambda]= \frac{2}{3}\rho\,\,,\qquad (\rm{UV},\,\rho \to \infty)
\label{re}
\eeq
This relation was proved in the unflavored backgrounds
by identifying the gaugino condensate (an operator of dimension 3)
with a deformation of the background that asymptotically behaves
as $e^{-2\rho}$, see \cite{Di Vecchia:2002ks,Apreda:2001qb}.
 Since in the flavored backgrounds, the same kind
of deformation (leading to the type {\bf N} solutions) still has the
same exponential behaviour \cite{Casero:2006pt},
(\ref{re}) is trustable in the present setup.

We now proceed doing the computation (\ref{betain}) in the UV (for 
$\rho\to\infty$). 
After substituting (\ref{esta}), (\ref{re})  and (\ref{newbps1}),
(\ref{newbps2}), one
obtains an expression for the anomalous dimension of the quark superfield 
$\gamma_Q$ in terms of the  background functions,
\be
\gamma_Q=\frac{12}{N_f} Y -\frac{1}{2} -3\frac{N_c}{N_f}
\,\,,\qquad (\rm{UV},\,\rho \to \infty)
\label{anomalous}
\ee
Inserting (\ref{UVsmallNf}), we find the UV behaviour of
the anomalous dimension for 
$N_f<2N_c$:
\beq
\gamma_Q\sim -\frac{1}{2} -\frac{12  N_c}{32 (2 N_c-N_f) \rho^2}
\,\,,\qquad (N_f<2N_c )
\label{anomalous1}
\eeq
This expression is quite nice, because it matches the fact that 
 the anomalous dimensions of the quark superfield 
approaches $\gamma_Q=-\frac{1}{2}$ to have the R-charge 
$R[Q]=1/2$, see the discussion in section \ref{sec:globmatch}.

One may wonder what happens in theories  with $N_f>2N_c$
(which are Seiberg dual to those with $N_c < N_f< 2N_c$).
Introducing in (\ref{anomalous})
 the expansion of $Y$ for $N_f>2N_c$, eq. (\ref{UVbigNf}), we find:
\beq
\gamma_Q\sim \frac{5}{2} -6\frac{N_c}{N_f} -\frac{3  (N_f-N_c)}{8 ( 
N_f-2N_c) \rho^2}\,\,,\qquad (N_f>2N_c )
\label{gammasei}
\eeq
So $\gamma_Q > -\frac12$, as anticipated in section \ref{sec: beta}.

\subsection{The quartic coupling}
We will now analyze similar issues for the quartic coupling in the superpotential of
(\ref{ourtheory}). Let us start with the beta function.
\subsubsection{Beta function of the quartic coupling}
\label{sec: bquartic}
Let us go back to the field theory expression of the beta 
functions as we discussed them in eq.(\ref{betas}). 
We have, for the (dimensionless) quartic coupling:
\beq
\beta_{\tilde\kappa} = \frac{\partial\tilde\kappa}{\partial 
\log(\mu/\Lambda)} 
=\tilde{\kappa} (1 + 2 \gamma_Q)
\label{betakappa}
\eeq
Using the radius-energy relation (\ref{re}) and the
value of the anomalous dimension computed in (\ref{anomalous}),
we find:
\be
\frac {\partial \log \tilde \kappa}{\partial \rho} = 
\frac23 ( 1 + 2 \gamma_Q) =
\frac{16 Y}{N_f} - 4 \frac{N_c}{N_f}
\,\,,\qquad (\rm{UV},\,\rho \to \infty)
\ee
So, in view of (\ref{newbps2}), we can identify:
\be
\log \tilde \kappa = c + \frac{8G}{N_f}
\,\,,\qquad (\rm{UV},\,\rho \to \infty)
\label{Uvquartic}
\ee
where $c$ is some constant.

Let us analyze what happens with the quartic coupling 
when $N_f < 2N_c$ and $N_f >2 N_c$.
Whether it is relevant or irrelevant depends on the sign of 
$1 + 2 \gamma_Q = \frac {6}{N_f} (4Y - N_c)$.

When $N_f < 2 N_c$:
\be
1 + 2 \gamma_Q = -\frac {2N_c}{3 (2N_c - N_f)} \rho^{-2} < 0
\ee
so the quartic coupling is relevant: 
starts at some constant in the UV and grows towards the IR.

When $N_f > 2N_c$,
\be
1 + 2 \gamma_Q = \frac{6}{N_f} (N_f - 2N_c)  > 0
\label{irrelevantquartic}\ee
so the quartic is irrelevant and diverges in the UV.
Notice the consistency of these results with the 
field theory discussion in \cite{Strassler:2005qs,Strassler:2003qg}.

\subsubsection{Geometric interpretation}

In principle, one could expect that the quartic coupling could be
read as data of the geometry, {\it i.e.} as some quotient
of the volumes of the different cycles of the manifold.
In that sense, (\ref{Uvquartic}) is not very satisfactory.

Without a definite understanding, we speculate in this section about a 
{\it definition} of the quartic coupling that makes use of the
geometry of the internal manifold. Since (\ref{Uvquartic})
is trustable in the UV, we require that our guess behaves
similarly in the large $\rho$ region.
We start with the $N_f < 2N_c$ in which the coupling is 
relevant and bounded
in the UV:
\bea
\tilde \kappa = e^{c+ \frac{8G}{N_f}} \approx
1 + \frac{N_c}{2(2N_c-N_f)} \rho^{-1}+\dots \,\,,\qquad
(N_f < 2N_c)
\eea
where we have conveniently fixed $c$ (an overall factor) to make
the expression simpler.

We will now propose a definition for the quartic coupling 
$\tilde{\kappa}$ in terms of the geometry. First, 
let us notice that, for fixed $\rho$, we have one well defined
two-cycle, a well defined one-cycle and a couple of well defined 
three-cycles. The 
two-cycle is the one mentioned before and given by
\beq
\hat{S}^2=[\theta=\tilde\theta,\;\; \varphi= 2\pi-\tilde\varphi],\;\;\; 
Vol[\hat{S}^2]= 4\pi e^{\phi}(H+G).
\label{dosciclo}
\eeq
The one-cycle, along the worldvolume of the flavor branes,
is parametrized by the 
angle $\psi$ and  its volume is
\beq
S^1=[\psi],\;\; Vol[S^1]= 4\pi e^{\phi/2}\sqrt{Y}.
\eeq
It is possible to define two different 3-cycles in the 
internal manifold
\beq
 S^3=(\theta,\varphi,\psi),\; Vol[S^3]=(4\pi)^2 
e^{3/2\phi}H\sqrt{Y},\;\;
\tilde{S}^3=(\tilde\theta,\tilde\varphi,\psi),\;\; 
Vol[\tilde{S}^3]=(4\pi)^2e^{3/2\phi} 
G\sqrt{Y}
\eeq
We {\it define} the quartic 
coupling 
as a quotient of the volume of the two-cycle times the one-cycle 
divided by the volume of one of the three-cycles. This definition makes 
sense since the coupling we are interested in (relevant for matchings 
between string and gauge theory) must be {\it adimensional}:
\beq
\tilde \kappa=\frac{Vol[S^1] \times Vol[\hat{S}^2]}{Vol[S^3]}
\label{definitionq}
\eeq
Then:
\beq
\tilde \kappa=1 + \frac{G}{H} = 1 +  
\frac{N_c}{2(2N_c-N_f)} \rho^{-1}+\dots
\label{quarticelect}
\eeq
When studying the Seiberg dual magnetic theory, one should
change $S^3 \to \tilde S^3$ according to the prescription
(\ref{cambi}). This amounts to the interchange 
$H \leftrightarrow G$.

When $N_f > 2N_c$, the quartic coupling grows exponentially
towards the UV and the definition (\ref{definitionq}) should be taken 
carefully. This is perhaps not surprising since the field theory
description with a quartic term breaks down and needs some
UV completion. We relegate further discussion to 
Appendix \ref{appendix2}.

\subsection{A nice matching with the $N_f=2N_c$ case}
In this section, we will show a connection between our solutions for
$N_f<2N_c$ with the one for $N_f=2N_c$ and  the one for $N_f>2N_c$,
and analyse it in view of the properties of the quartic
coupling discussed above.
We will need to use that all of the solutions are of the form
(\ref{metricA}), with the expansions in 
eqs.(\ref{UVsmallNf})-(\ref{UVbigNf})
and that for $N_f=2N_c$ an explicit solution is given by  
\cite{Casero:2006pt} ($\xi$ is a free parameter valued in $0<\xi<4$),
\beq
H(\rho)=\frac{N_c}{\xi},\;\;\; G(\rho)=\frac{N_c}{4-\xi},\;\;\; Y=
\frac{N_c}{4},\;\;\; \phi(\rho)= \phi_0 +\rho .
\label{nfi2nc}
\eeq
We now consider the solution with
$N_f<2N_c$ in the far UV ($\rho\to\infty$), 
and change the number of flavors towards $N_f \to 2N_c^-$
with a particular  
scaling limit
\beq
\lim_{\rho\to\infty,\;\; N_f\to 2N_{c}^-} (2N_c-N_f)\rho\to \infty ,
\eeq
so we have that the functions in eq.(\ref{UVsmallNf}) behave as
\beq
H\approx \Big(\frac{2N_c - N_f}{2}\Big) \rho\to \infty,\;\; G\approx
\frac{N_c}{4},\;\; \phi\approx \rho.
\eeq
According to (\ref{Uvquartic}), the quartic coupling takes some
finite value at this point.

We see that we can match this with the $N_f=2N_c$ solution (\ref{nfi2nc})
if we take $\xi= 0$. 
After matching, we move freely in the parameter
$\xi$ 
(since in the $N_f=2N_c$ case, $\xi$
represents a marginal coupling \cite{Casero:2006pt}), from $\xi=0$ towards
$\xi=4$.
At the point $\xi=4$ the functions take the values
\beq
G\to \infty,\;\; H=\frac{N_c}{4},\;\; \phi=\rho
\label{zz1}\eeq
and the quartic coupling diverges.
We can match (\ref{zz1})
with the expansion for $N_f > 2N_c$ in eq.(\ref{UVbigNf}) if we scale
similarly as above,
\beq
\lim_{\rho\to\infty,\;\; N_f\to 2N_{c}^+} (N_f-2N_c)\rho\to \infty.
\eeq

The interpretation is the following: since when $N_f > 2N_c$, the
quartic is irrelevant, we can only connect those theories with the
end of the $N_f = 2N_c$ marginal line in which this coupling is
infinite. On the other hand, when $N_f < 2N_c$, the
quartic is relevant so these theories approach the conformal
$N_f = 2N_c$ point by the opposite limit of the marginal line, where
the quartic takes its smallest value. At $N_f = 2N_c$, we can tune
$\xi$ (accordingly changing the quartic coupling)
to move from one end of the line to the other. The process
continuously connects, in the space of theories,
 the $N_f < 2N_c$ case to the $N_f > 2N_c$ one.

This ``flow in  the
number of flavors'' provides a nice quantitative check and suggests that
our discussion about the quartic coupling in section
\ref{sec: bquartic}
is sensible.
It would be nice to understand if  similar matchings and flows may
happen for different solutions with $N_f=2N_c$ \cite{Caceres:2007mu}.

Let us now study the potential between static quark-antiquark and 
monopole-antimonopole pairs, using Wilson and 't Hooft loops.

\subsection{Wilson Loop and QCD-string}
\label{sec: wil}
In order to get some intuition of what will happen to the quark-antiquark 
pair when separated, we can  analyze a quantity
associated with the tension of the putative QCD-string that 
forms between them. The functional form of this tension (that can be 
thought of as the tension of an F1-string when taken to the IR of the 
background) is given by,
\bea
& & T_{Q\bar{Q}}\sim \sqrt{g_{tt}g_{xx}}|_{IR}= e^{\phi}|_{IR}\nonumber\\
& & T_{Q\bar{Q},{Type I}}\to 0\nonumber\\
& & T_{Q\bar{Q},{Type II}}\to e^{\phi_0}\nonumber\\
& & T_{Q\bar{Q},{Type III}}\to e^{\phi_0}
\label{qcdstringtension}
\eea
$T_{Q\bar{Q}}$ gives a {\it qualitative} way of evaluating the tension in 
the IR dynamics.
The expressions (\ref{qcdstringtension})
 tell us that type I solutions will have vanishing QCD-string tension 
(indicating that quarks are free), type II and type III solutions will show 
finite tension, 
indicating confinement. 
This is in agreement with the behavior of the 
gauge coupling in eq.(\ref{esta}). Of course, the presence of dynamical 
quarks will produce screening (string breaking) when the energy stored in 
the QCD-string (\ref{qcdstringtension}) is of the order of the lightest 
meson in the theory.

Let us now compute the Wilson loop following 
 the standard prescription 
of \cite{Maldacena:1998im}.
Consider the action for an F1 string  parametrized 
by
\beq
t=\tau,\;\;x=\sigma,\;\; \rho=\rho(\sigma)
\label{cwl}
\eeq
with $x$ taking values in 
$[0,L]$ (such that $\rho(0)=\rho(L)=\infty$)
and $\tau$ in $[0,\infty)$.
The induced metric on this string is,
\beq
ds^2_{ind}= e^\phi \Big[-d\tau^2 +
(1+4 Y {\rho'}^2) d\sigma^2\Big]
\label{indwilson}
\eeq
So, the Nambu-Goto action for this string is
\beq
S=\frac{1}{2\pi\alpha'} \int d\sigma e^{\phi}\sqrt{1 + 4Y {\rho'}^2}
\label{ang}
\eeq
This is in the form to follow the usual treatment for Wilson loops 
in \cite{Sonnenschein:1999if}, 
and write expressions for the (renormalized)
energy $E$ and 
the 
separation $L$.
There is a one-parameter family of string profiles depending
on the minimal value of $\rho$ reached by the string, which we call
$\rho_0$.
\bea
& & E (\rho_0)
= \frac{1}{2\pi\alpha'} \left[
e^{\phi(\rho_0)} L + 2 \int_{\rho_0}^\infty 2\sqrt{Y}(\sqrt{e^{2\phi} - 
e^{2\phi(\rho_0)} } - e^{\phi})d\rho - 2 \int_0^{\rho_0} 2 
\sqrt{Y}e^\phi  d\rho \right] \nonumber\\
& & L (\rho_0)
= 4 e^{\phi(\rho_0)}\int_{\rho_0}^\infty \sqrt{\frac{Y}{e^{2\phi} -
 e^{2\phi(\rho_0)}}}d\rho
\label{EL}
\eea
These integrals can be studied numerically for  
our solutions. We will present an example in 
figure \ref{fig: loops} at the end of the next section.

\subsection{t'Hooft loop}
\label{sec: tho}

It is interesting to perform a qualitative analysis 
similar to the one around eq.(\ref{qcdstringtension}). In this case, the 
effective tension for the monopole-antimonopole pair is given by the 
effective tension of the magnetic string, 
represented by a D3-brane wrapping
the two-cycle (\ref{dosciclo}) and extended along $(x,t)$ stretched at the bottom of the geometry.
See eq.(\ref{cuatrocycleee}) and below for details,
\bea
& & T_{m\bar{m}}\sim e^{\phi} (H+G)|_{IR}\nonumber\\
& & T_{m\bar{m},{Type I}}\to e^{\phi_0}N_f\nonumber\\
& & T_{m\bar{m},{Type II}}\to e^{\phi_0} C\nonumber\\
& & T_{m\bar{m},{Type III}}\to 0
\label{mmstringtension}
\eea
$T_{m\bar{m}}$ gives a {\it qualitative} way of evaluating the tension of 
the monopole-antimonopole pair in the IR dynamics
Intuitively, eq. (\ref{mmstringtension}) 
indicates that type I and type II solutions 
confine the monopoles, while type III solutions exhibit free monopoles.

\begin{figure}[htbp] %  figure placement: here, top, bottom, or page
  \centering
  \includegraphics[width=0.45\textwidth]{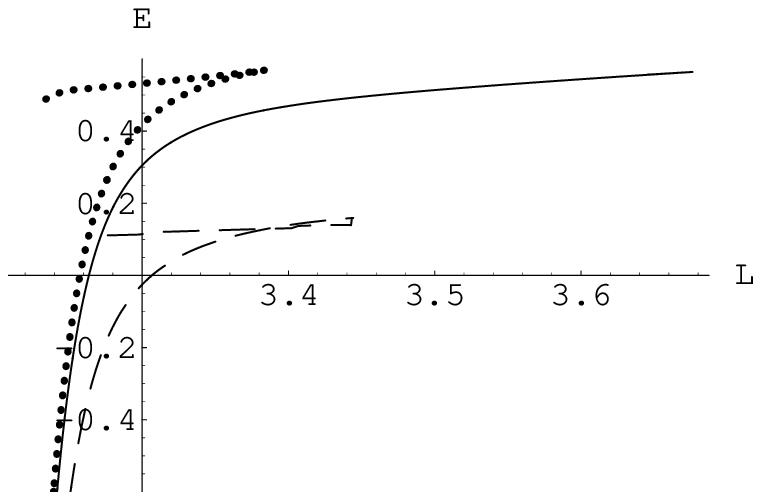}\hfill
  \includegraphics[width=0.45\textwidth]{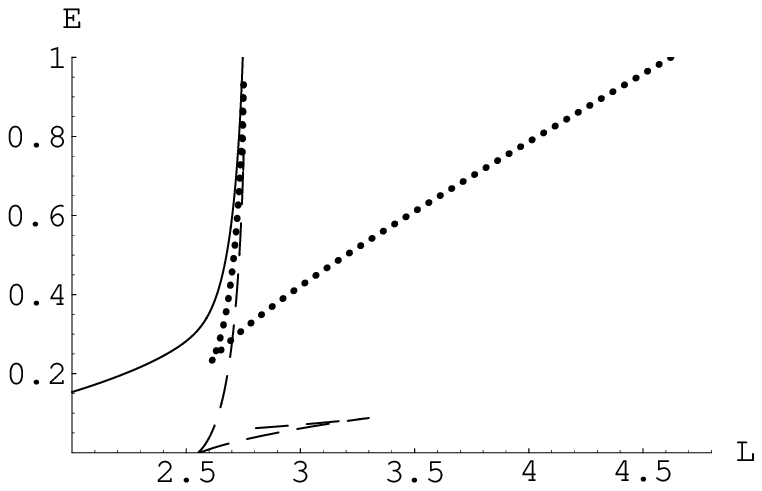}
  \caption{As a typical example, we plot   E vs. L graphs
for $N_f = 1.2 N_c$. The dotted line corresponds
to the type I IR behaviour, the dashed line to type II (with $C=-1$) and the solid line
to type III. The graph in the left is for Wilson loops as discussed in section \ref{sec: wil} while
the one in the right corresponds to the 't Hooft loops introduced in section \ref{sec: tho}}
  \label{fig: loops}
\end{figure}

Let us now study in detail
the t' Hooft loop that can be thought of as the 
``Wilson loop for magnetic monopoles''(the rigorous 
definition is in terms of the 
commutation relations with the Wilson loop operator). 
It is a similar computation to the one
for the Wilson loop in the 
previous section, as we explained above, for a extended wrapped D3 
brane on the cycle $\Sigma_4$,
\beq
\Sigma_4=\Big(t=\tau,\;\; x_1=\sigma,\;\;\; \rho(\sigma),\;\; 
\theta=\tilde{\theta}, 
\;\;\varphi=2\pi- \tilde{\varphi}\Big).
\label{cuatrocycleee}\eeq
This gives an induced metric that in string frame reads,
\beq
ds_{ind}^2= e^{\phi}\Big[ -dt^2   +(1+ 4 Y {\rho'}^2 ) d\sigma^2 + 
(H+G)(d\theta^2 +\sin^2\theta d\varphi^2)\Big]
\label{indd3th}
\eeq
and the action (per unit time) for the `effective string' obtained after 
integrating out 
the two-cycle is,
\beq
S= 4\pi T_3\int d\sigma \sqrt{\hat{f}^2 + 
\hat{g}^2{\rho'}^2},\;\;\; \hat{f}= e^{\phi}(H+G),\;\;\; \hat{g}= 2\hat{f} 
\sqrt{ Y}
\label{thooftloop}
\eeq
We observe that this expression is ``Seiberg invariant'' according to 
the definition (\ref{seiberg}), and  analogous formulas 
to those in eqs. (\ref{EL}) can be used. They read
\bea
& & L (\rho_0) =2\int_{\rho_0}^{\infty} 
\frac{\hat{g}}{\hat{f}}\frac{\hat{f}(\rho_0)}{\sqrt{\hat{f}^2(\rho) - 
\hat{f}^2(\rho_0)}} d\rho\quad\nonumber\\
& & E=4\pi T_3 \left[\hat{f}(\rho_0) L + 2\int_{\rho_0}^{\infty} 
\frac{\hat{g}(\rho)}{\hat{f}(\rho)}\Big({\sqrt{\hat{f}^2(\rho) -
\hat{f}^2(s_0)}} - \hat{f}(\rho)  \Big)d\rho -2 \int_{0}^{\rho_0} 
\hat{g}(\rho) d\rho\right]\quad
\label{EL2}
\eea
An example of numerical analysis for the expressions
 (\ref{EL2}) using the functions (\ref{thooftloop}) is plotted
in figure \ref{fig: loops}.

Our interpretation of figure \ref{fig: loops} is as follows: first, it is interesting to
notice that the graphs for the backgrounds with different IR behaviours
share some qualitative properties, suggesting a smooth interpolation among
the different solutions. In all cases, there is a maximum length reached by
the loop which we interpret as string breaking due to pair creation.
\footnote{
In \cite{Apreda:2003sy}, it was argued in a similar
situation that such kind of behaviour did not correspond to string breaking.}
Moreover, in both graphs, there is a line 
(the type III one for the Wilson loop and the one corresponding to
type I for the 't Hooft loop)
which stretches more than the rest indicating linear confinement (before the string
is eventually broken). This qualitatively suggests a change of roles of
quarks and monopoles between type I and type III backgrounds. We will 
elaborate on this in the next section.
Finally, let us point out that there are also minimal lengths in the graphs. 
Presumably, this is due to the growing linear dilaton, which signals the
failure of the supergravity description we are using at asymptotically large $\rho$.
With such an UV completion, there should be a minimal length ($L_{min}$) of the field theory that can
be reliably probed by the present 
computation.
Indeed, the converging growing lines on the left of the 't Hooft loop graph
are an artifact of this UV behaviour (length and energy in equations (\ref{EL2}) grow with $\rho_0$
when $\rho_0$ is taken large). Thus, we discard this part 
of such lines as unphysical.
 The fact that the type III line of the 't Hooft loop graph
only follows this unphysical behaviour indicates that in 
the corresponding
field theory, monopoles are free (or they are screened at 
a length shorter than $L_{min}$).

We close this section with a technical comment about the obtention of the graphs in figure
\ref{fig: loops}. In order to compare the loops probing the different solutions, one has
to ensure that they asymptote to the same UV (once we fix the additive constant of integration
$\phi_0$ for one of the solutions, the $\phi_0$ of the rest have to be fixed accordingly).
It is also necessary to take into account that having $C\neq 0$, the UV behaviour of
$H$ is different from the $C=0$ case (see eq(\ref{UVsmallNf})). Thus, in order to
compare with other solutions, the definition of the coordinate $\rho$ has to be shifted
$\rho \to \rho - \frac{2C}{2N_c - N_f}$ (so the origin of space is not at $\rho=0$ any more).

\subsection{Back to the 1970's}
\label{backto1970}
Briefly, we want to make contact with the material in section 
\ref{jewel1970}. 

The intuitive analysis for the string tensions in eq.(\ref{qcdstringtension}) 
and eq.(\ref{mmstringtension}), shows that in the three types of IR 
solutions the behavior for the potential between the (non-dynamical) 
quark-antiquark and monopole-antimonopole 
pair is quite diverse. Indeed, the intuitive analysis presented around 
eqs.(\ref{qcdstringtension})-(\ref{mmstringtension}) shows that for 
small separations compared to the mass of a pair, type I solutions confine 
monopoles and let quarks move 
freely. The opposite  occurs in solutions of type III, while 
type II solutions present a situation in which quarks and monopoles 
experience a growing force when separated \footnote{We emphasize again 
that for long distances 
pair creation gives place to string breaking and the law changes.}. 

Another important observation is the fact that one can smoothly 
interpolate 
between the  solutions of types I, II and III by changing the values of 
the integration 
constants. Indeed, one can see that in the limit of vanishing constant $C$ 
in type II solutions (see section \ref{typeiiexpansionss}), the 
expansions change into those of type III in section 
\ref{typeiiiexpansionss} (this was repeatedly stressed when plotting 
the functions). 
The connection with type I solutions in section \ref{typeiexpansionss}
must be thought in a more heuristic way, by noticing 
that if the constants $C$ and $h_1$ in the type II solutions 
 were absent the expansions could continue past $\rho=0$ to 
smaller values of the radial coordinate and the leading term in the 
expansion would go like the type I solutions (this
connection could be made more precise in terms of the UV constants
of integration $C,Q_{1,0}$, see section \ref{sec: revisit}, but this is
hard to see in the numerical analysis).

This `heuristic matching' resembles the fact that 
in theories with fundamental fields there are no phase transitions in moving 
between a confining and a higgsed situation \cite{Fradkin:1978dv}. The 
different VEV's of the
operator of dimension six mentioned in section \ref{sec: revisit} is what 
should take the theory from one vacua (a particular string background) to 
another.
Together with the behavior for Wilson and 't Hooft loops, 
this leads us to propose
that the solutions of type I, II and III behave 
like the Higgs, oblique 
confined and confined regimes of the SQCD-like theory in 
eq.(\ref{ourtheory}).
In the oblique confined phase, some dyons condense and
become free. This should be achieved in the present setup by
the vanishing tension of 
 a wrapped extended D3-brane with some F-strings disolved 
inside it, {\it i.e.} with the worldvolume electric field 
turned on. We leave this computation for the future.

The qualitative picture described above is quite appealing.
However, we should write here a word of caution. The
tensions (\ref{qcdstringtension}) and (\ref{mmstringtension})
depend on a combination of functions at the singularity; 
even when the singular behavior cancels from the expressions of 
(\ref{qcdstringtension}) and (\ref{mmstringtension}) -a 
manifestation of the `good' character of the singularity-
these results should be taken with care.
It could happen that the criterion we are using for
`good singularities' is not enough
and that not all the discussed  solutions should be accepted as physical.

None of the described behaviours corresponds to an IR conformal fixed
point. We remind the reader that SQCD with a quartic superpotential
flows in the IR to Seiberg's conformal fixed point of SQCD
\cite{Strassler:2005qs} (at least in the ratio $N_f/N_c$ appropriate to 
fall in the conformal window). We believe that this does not show up in our
set-up since, as explained in section \ref{nfbignc}, the theory dual to the
string solution differs from the usual SQCD+quartic near the IR, presumably
due to a non-trivial VEV.

\subsection{A comment on the running couplings}\label{morefield}
Coming back to the expressions for the Wilson and 't Hooft loops in 
eqs.(\ref{ang})-(\ref{thooftloop}), we observe that they are invariant 
under the Seiberg duality transformation (\ref{seiberg}). Even more, one 
can easily see that the gauge coupling defined in eq.(\ref{esta}) is also 
invariant. This may puzzle the reader, who may expect that while the 
electric theory is strongly coupled in the IR, the magnetic theory should be 
IR free and vice versa. 
This is not what we observe; our solutions predict that both theories are 
strongly coupled (or both have weak $g_{YM}^2$ coupling) towards the IR.
This situation is not totally original, it is indeed what 
happens inside the conformal window in Seiberg's SQCD. In contrast, in our 
case, we do not have a conformal window, but we will see that having two 
theories with similar IR behavior is not ruled out.

Let us discuss this issue using the expressions for the 
wilsonian beta function in eq.(\ref{betain}) and the values for the 
anomalous 
dimensions we obtained in 
eqs.(\ref{anomalous1})-(\ref{gammasei}) for the $N_f<2N_c$ and $N_f>2N_c$ cases 
respectively.
First, we introduce the  notation: the theory $\bf{e}$ is the one 
with gauge group $SU(N_c)$ and $N_f$ flavors. 
We will assume that in the $\bf{e}$ theory $N_f<2N_c$,
and the superpotential is,
\beq
{\cal{W}}_e=\kappa_e \Big(\tilde{Q} Q \tilde{Q} Q -\frac{1}{N_c} 
(\tilde{Q} 
Q)^2 \Big)
\label{superpot4}
\eeq
We will denote by {\bf m} theory the one obtained by Seiberg duality 
from the theory {\bf e}. The {\bf m}~theory will have gauge group $SU(n_c)$ 
with $N_f$ flavors, of course the relation is $n_c=N_f-N_c$. This, 
together with the fact that in the {\bf e} theory we set $N_f< 2N_c$ 
implies that $N_f>2 n_c$ in the theory {\bf m}. The superpotential in the 
{\bf m} 
theory is 
(we will ignore, since it plays no role, the double trace term),
\beq
{\cal{W}}_m= \hat{\kappa}_m M M +\frac{1}{\mu}\tilde{q}M q
\label{superpot4b}
\eeq
where we have used that the new fields in the Seiberg dual theory are the 
meson $M_{ij}$ (not charged under the $SU(n_c)$ gauge field $w_\alpha$), 
the quarks $q$ and the antiquarks $\tilde{q}$.

Now, let us compute the beta functions in each 
theory. For this, we will need to know the anomalous dimensions of the 
fields. Our solutions 
predict (in the UV) anomalous dimensions for the {\bf e} quarks $\gamma_Q$ 
and the {\bf 
m} quarks $\gamma_q$ given by eqs.(\ref{anomalous1}) and 
(\ref{gammasei}) 
\bea
& & \gamma_Q=\frac{12}{N_f}Y -\frac{1}{2} -3\frac{N_c}{N_f}\sim -\frac{1}{2} 
-\frac{12  N_c}{(64N_c-32N_f)\rho^2}\nonumber\\
& & \gamma_q=\frac{12}{N_f}Y -\frac{1}{2} -3\frac{n_c}{N_f}\sim \frac{5}{2} 
-6\frac{n_c}{N_f}-\frac{3(N_f-n_c)}{(8N_f-16n_c)\rho^2}.
\label{anomalousuv}
\eea 
Let us neglect the small $O(\rho^{-2})$ correction in the far UV 
($\rho\to\infty$ region), 
keeping 
only the leading terms.
Now, we compute the beta functions. We use the values for the 
anomalous dimensions in eq.(\ref{anomalousuv}), obtaining that the 
wilsonian beta functions for the two theories are given by,
\bea
& & 
\beta_{\frac{8\pi^2}{g_e^2}}= 3 N_c - 
N_f(1-\gamma_Q) =\frac{3}{2}(2N_c-N_f),\nonumber\\
& & \beta_{\frac{8\pi^2}{g_m^2}}=3 n_c - N_f(1-\gamma_q)= 
\frac{3}{2}(N_f-2n_c),\nonumber\\
\label{betaszz}
\eea
This shows that either both the electric and magnetic theories confine or they both are IR 
free. The result in eq.(\ref{betaszz}) is precisely the one we obtain if 
we use the UV expansions for $H(\rho), G(\rho)$ in section \ref{sec: UV} 
together with eqs.(\ref{anomalous})-(\ref{gammasei}). We now move to the 
matching of anomalies.

\subsection{Global Anomaly matching}
\label{sec:globmatch}
In a field theory like SQCD or 
ours in eq.(\ref{ourtheory}) there are two types of currents, global 
(denoted below by $j_g$) and local or gauge currents, denoted by $J_L$.
Hence, there are four possible 3-point correlators of currents one 
may compute
\bea
& & <J_L J_L J_L>,\label{anomagau}\\
& & <j_g j_g J_L>,\label{anoma}\\
& & <j_g J_L J_L>,\label{anomalous3pt}\\
& & <j_g j_g j_g>.\;\; \label{thoft}
\eea
The first and second correlators (\ref{anomagau})-(\ref{anoma}) must be 
zero, or the gauge symmetry is anomalous and the theory is inconsistent.  

The correlator with two local currents in 
eq.(\ref{anomalous3pt}) if nonzero, indicates 
the presence of an anomaly in the global current $<\partial_\mu j^\mu_g >\neq 0$. 
While for non-anomalous symmetries in the sense of 
eq.(\ref{anomalous3pt}), the correlator for three global currents in 
eq.(\ref{thoft})  must  match in the two descriptions of the theory. This 
is the 't Hooft anomaly matching condition.

We will start by studying the anomaly for the $R$ symmetry current, as 
expressed by the  correlator in (\ref{anomalous3pt}), and related to the
results of sections \ref{anomalymatching} and  \ref{sec:thetagym}.
This is what appears in the changes for the $\Theta$ angle. A general 
expression for this anomaly is,
\beq
\psi\to e^{i\epsilon\gamma_5}\psi,\;\;\;\Delta \Theta= \epsilon\sum_r n_r T(r)
R[\psi]
\label{psitransf}
\eeq
where the $\psi$ denotes the massless fermions of the theory,
$n_r$ is the number of particles in the $r$ representation of the
gauge group, $T(r)$ is the weight of the representation (we remind the
reader that
$T(adj)=2N_c,\;T(fund)=1$) and $R[\psi]$ is the R-charge of the
corresponding fermion. 
We need to 
assign R-symmetry charges to the superfields. The gaugino has 
$R[\lambda]= 1$ and we want to determine the R-charges of the fundamental
superfields. They cannot be read directly from the supergravity solution,
but they can be extracted indirectly from the $\Theta$-angle transformation.
Let us start looking at the electric {\bf e}-theory (with $N_f < 2N_c$). From
(\ref{psitransf}), we read (using $R[\psi_Q]= R[\psi_{\tilde{Q}}]$, required by the symmetry):
\be \Delta \Theta_e=\epsilon_e (2N_c + 2 N_f R[\psi_Q])
\ee
 The values of $\epsilon_e$ that leave $\Theta_e$ invariant (modulo $2\pi$)
are written in (\ref{match}) and immediately suggest $R[\psi_Q]=-\frac12$.
Thus, we are led the following assignation (for the 
fermions in the multiplet) under the global symmetry 
$SU(N_f)_V \times U(1)_B \times U(1)_R$,  for the electric {\bf e}-theory:
\beq
\psi_Q=(N_f, 1, -1/2),\;\; \psi_{\tilde{Q}}=(\bar{N}_f, -1,- 
1/2),\;\;\lambda=(1, 0, 
1)
\label{valuescargas}
\eeq
This implies that for the scalars of the quark multiplets
$R[Q]=R[\tilde{Q}]=\frac{1}{2}$. Notice that this is consistent with the
existence of the quartic superpotential preserving the R-symmetry at the
classical level. Let us also discuss the relation of this R-charge to the
anomalous dimension of the quark field. At a conformal point
\beq
dim O=\frac{3}{2}R[O]
\label{conformalpointrelation}
\eeq
Using this relation and the UV value of the anomalous dimension
of the squark (\ref{anomalous1}), we find
$R[Q]= \frac23 dim [Q] = \frac23 (1 + \frac12 \gamma_Q)=\frac12$,
in agreement with the discussion above.
The relation (\ref{conformalpointrelation}) can be
used here because our theory has an asymptotic UV fixed point
(the beta-function for the
gauge coupling becomes trivially zero since the gauge 
coupling itself goes to zero and the one for the quartic coupling
also vanishes due to $\gamma_Q=-\frac{1}{2}$, see (\ref{betakappa})).

We now turn to  the dual magnetic {\bf m}-theory with $n_c= N_f - N_c$
colors such that $N_f > 2n_c$. We denote the magnetic quarks 
with lower case $q$. In this case,
(\ref{psitransf}) yields:
\be
\Delta \Theta_m=\epsilon_m(2n_c + 2 N_f R[\psi_q] )
\label{magtheta}
\ee
Comparison to  (\ref{match}) again suggests $R[\psi_q]=-\frac12$, which
we believe is the charge relevant to the backgrounds discussed in this paper.
However, there is another logical possibility which we find interesting
and we will discuss it in section \ref{sec:othercharges}.
Therefore, we consider the following charges   for the magnetic {\bf m}-theory:
\bea
& & \psi_q=(\bar{N}_f, \frac{N_c}{n_c}, -\frac{1}{2}),\;\;\psi_{\tilde{q}}=(N_f, 
-\frac{N_c}{n_c}, -\frac{1}{2})\nonumber\\
& & \psi_M=(N_f \times \bar N_f, 0, 0),\;\;\tilde{\lambda}=(1, 0, 
1)
\label{valuescharges}
\eea
This implies that for the scalars of the quark multiplets
$R[q]=R[\tilde{q}]=\frac{1}{2}$ and for the meson multiplet $R[M]=1$, again
allowing the superpotential to preserve the R-symmetry.
However, given (\ref{gammasei}), one can see that the relation (\ref{conformalpointrelation})
is not satisfied.
This should not come as a 
surprise, because the quartic coupling grows in the UV of the magnetic 
theory, see around eq.(\ref{irrelevantquartic}) and the UV theory is far 
from a conformal point.

Indeed, now that the assignations in 
eqs.(\ref{valuescargas}),(\ref{valuescharges}) have been given, the next 
step
is to check whether 't Hooft matching in the sense of eq.(\ref{thoft}) 
holds. Since the R-symmetry is anomalous, the continuous global
symmetry  is $SU(N_f)_V \times U(1)_B $.
The triangles that should be matched are,
\beq
<SU(N_f)_V^3>,\;\; <SU(N_f)_V^2 U(1)_B>, \;\;<U(1)_B^3>, \;\; <U(1)_B TT> 
\eeq
where the last triangle representes the `gravitational anomaly' with 
two insertions of $T_{\mu\nu}$. Let us quote the results in the {\bf e} 
theory with the values of eq.(\ref{valuescargas}), we denote 
$d^{ABC}=Tr(T^A [T^B,T^C]) $,
\bea
& & <SU(N_f)_V^3> = d^{ABC} N_f N_c (1^3 + (-1)^3) =0\nonumber\\
& & <SU(N_f)_V^2 U(1)_B> = tr(T^A T^B) 2 N_f N_c (1-1)=0\nonumber\\
& & <U(1)_B^3> =N_f N_c (1^3 + (-1)^3)=0\nonumber\\
& & <U(1)_B TT> = N_f N_c (1-1)=0
\label{thoftxxx}
\eea
while for the {\bf m} theory, with values (\ref{valuescharges}), we have
\bea
& & <SU(N_f)_V^3> = d^{ABC} N_f n_c (1^3 + (-1)^3) =0\nonumber\\
& & <SU(N_f)_V^2 U(1)_B>= tr(T^A T^B) 2 N_f n_c 
(\frac{N_c}{n_c}-\frac{N_c}{n_c})=0\nonumber\\
& & <U(1)_B^3> =N_f n_c [(\frac{N_c}{n_c})^3 + 
(-\frac{N_c}{n_c})^3]=0\nonumber\\
& & <U(1)_B TT> = N_f n_c (\frac{N_c}{n_c} - \frac{N_c}{n_c})=0
\label{thoftzzz}
\eea
We see that there is anomaly matching between the electric and 
magnetic descriptions, satisfying 't Hooft criteria.
Up to here, the computation is identical to that for the usual 
${\cal N}=1$ SQCD without superpotential.
In \cite{Csaki:1997aw}, it was shown in  that triangles involving 
discrete symmetries should also be matched. It is therefore a 
non-trivial consistency check to compute the values
involving the remnant non-anomalous R-symmetry $Z_{2N_c-N_f}=Z_{N_f-2n_c}$.

We now explicitly check the matching of {\it discrete} anomalies 
between the electric and magnetic description. Following the paper 
\cite{Csaki:1997aw}, we can compute the values for the triangles
involving the  R-symmetry. 
There is a technical subtlety explained in \cite{Csaki:1997aw}. We 
need to have integer values for the different charges to match discrete 
anomalies, so they have to be rescaled. Thus, following the prescription in
\cite{Csaki:1997aw}, we consider a $Z_k = Z_{4N_c - 2N_f}$ R-symmetry.
 The baryonic and R-charges are `rescaled'
 with respect to the values in the 
{\bf e}-theory eq.(\ref{valuescargas}) for the fermion 
in the multiplet and read:
\beq
\psi_Q=(N_f, \frac{N_f-N_c}{n}, -1),\;\; \psi_{\tilde{Q}}=(\bar{N}_f, -\frac{(N_f-N_c)}{n},- 1),
\;\;\lambda=(1, 0, 2)
\label{valuescargas3}
\eeq
while for the dual theory {\bf m}, rescaling in the same fashion the 
values in eq.(\ref{valuescharges}), we have
\bea
& & \psi_q=(\bar{N}_f, \frac{N_c}{n}, - {1}),\;\;
\psi_{\tilde{q}}=(N_f, -\frac{N_c}{n}, -{1})\nonumber\\
& & \psi_M=(N_f \times \bar N_f, 0, 0),\;\;\tilde{\lambda}=(1, 0, 2)
\label{valuescharges3}
\eea
The integer  $n$ is defined as the greatest common divisor of $N_c$ and $N_f$:
\be
n\equiv GCD (N_c, N_f)
\ee
One has to study the following triangles:
\beq
<Z_k TT>,\;\; <Z_k U(1)_B^2>,\;\; <Z_k^3>,\;\; <SU(N_f)^2 Z_k>
\label{kkk1}
\eeq
We will consider the difference between the triangles in the electric and 
the magnetic theory.
They should match up to multiples of $k=4N_c- 2N_f$ \cite{Csaki:1997aw}.
We obtain the following results,
\bea
& & <Z_kTT>_{e}=2 N_c N_f (-1)+2(N_c^2-1),\;\;<Z_kTT>_{m}=2 n_cN_f(-1) + 2 
(n_c^2-1),\nonumber\\
& &\Delta(e-m)=0
\eea
which is a multiple of $k=4N_c-2N_f$ as required by \cite{Csaki:1997aw}. 

For the second triangle involving three `discrete' currents,
\bea
& & <Z_k^3>_e= 2N_cN_f(-1)^3 +8(N_c^2-1),\;\;<Z_k^3>_m=2n_cN_f(-1)^3 
+8(n_c^2-1),\nonumber\\
& & \Delta(e-m)= 6N_f(2N_c-N_f).
\eea
Again, the difference is a multiple of $k$. For the triangle involving two 
baryonic and a `discrete' current,
\bea
& & <Z_k U(1)_B^2>_e= 2N_cN_f \left( \frac{N_f-N_c}{n}\right)^2 (-1),\;\; <Z_k 
U(1)_B^2>_m=2n_cN_f(-1)\left( \frac{N_c}{n}\right)^2,\nonumber\\
& &\Delta(e-m)=2\frac{N_cN_f}{n^2}(N_f 
-N_c)(2N_c-N_f),\eea
we obtain the correct matching modulo $k$; finally for the triangle 
involving the non-abelian global symmetry 
\bea
& & <SU(N_f)^2 Z_k>_e=  -2N_cN_f \;\;<SU(N_f)^2 Z_k>=-2n_cN_f   
\nonumber\\
& &\Delta(e-m)=-2 N_f(2N_c-N_f),
\label{discretematching}
\eea
We see that the assignation of charges 
 appropriately rescaled
(\ref{valuescargas3}), (\ref{valuescharges3})
satisfies the discrete anomaly matching up to multiples of $k=(4N_c-2N_f)$ 
as explained in \cite{Csaki:1997aw}. 
This nice match indicates that the value for the R-charge suggested by 
the string background is consistent with the 
field theory computations using the charges in eqs.
(\ref{valuescargas}),(\ref{valuescharges}).

\subsubsection{A different charge assignation}
\label{sec:othercharges}

We again consider the charges for the electric theory
written in (\ref{valuescargas}), but 
give different R-charges for the magnetic theory.
Coming back to (\ref{magtheta}), a second possibility
to match the supergravity result (\ref{match}) is to take
$R[\psi_q] = \frac12 -2 \frac{n_c}{N_f}$.

Our second assignation of charges for the magnetic theory
under the symmetries is, thus
(like before, we give the charge of the fermions in the multiplet):
\bea
& & \psi_q=(\bar{N}_f, \frac{N_c}{n_c}, 
\frac{1}{2}-2\frac{n_c}{N_f}),\;\;\psi_{\tilde{q}}=(N_f, -\frac{N_c}{n_c}, 
\frac{1}{2}-2\frac{n_c}{N_f})\nonumber\\
& & \psi_M=(N_f \times \bar N_f, 0, -2 + 4\frac{n_c}{N_f}),\;\;\tilde{\lambda}=(1, 0, 
1)
\label{valuescharges2}
\eea
The nice thing about this second assignation is that it satisfies the 
relation in eq.(\ref{conformalpointrelation}), as can be checked by using
 (\ref{gammasei}) 
(although, as explained above, there is no reason to expect
(\ref{conformalpointrelation}) to hold out of a conformal point). The downside is that this 
implies that the coupling $\hat{\kappa}_m$ in eq.(\ref{superpot4b}) must 
be charged, so the  R-symmetry is completely broken at the classical level in the
magnetic theory. This could only correspond to the dual of an
electric theory in which the R-symmetry is spontaneously broken and, thus,
it does not seem relevant to the type {\bf A} backgrounds discussed in this paper.
We believe that this second assignation could
describe a situation like 
what happens for type {\bf N} background in eq.(\ref{nonabmetric424}) or 
those solutions found in \cite{Casero:2006pt}, which will have the same UV 
as our type {\bf A} backgrounds.
Since in this case there is no remnant R-symmetry, an anomaly matching
computation
as the one in (\ref{kkk1})-(\ref{discretematching}) does not apply.

In summary, to match in field theory the R-symmetry anomalies to the ones 
computed by the string solutions, 
we proposed values for the charges of the multiplets 
under the global and discrete symmetries.
Notice that the R-charges are checked by the string background
in the sense that the anomaly is encoded in the background via
the potential $C_{(2)}$ (see section \ref{anomalymatching}). Both 
charge
assignations we have given are consistent with this computation.
On the other hand, the charges under $SU(N_f)$ and $U(1)_B$
written above are suggested by field theory considerations but
not from the actual string solution.
However, it is remarkable that the string solution indirectly knows
about them (at least
in the case of  assignation (\ref{valuescargas}),(\ref{valuescharges}))
throught the discrete anomaly matching computation
displayed in eqs. (\ref{kkk1})-(\ref{discretematching}).
The results in this section together with the 
previous checks, 
make us very confident of the interpretations that we have put forward 
for the string background.

\section{Summary, conclusions and future directions}
Let us briefly summarize the different topics elaborated in this paper. 
We have found a set of solutions for backgrounds of type {\bf A} in 
eq.(\ref{metricA}), proposed 
to be dual to a version of ${\cal N}=1$ SQCD (\ref{ourtheory}). With the new
solutions, we  predicted the behavior for different field 
theory observables and phenomena like Seiberg duality, gauge coupling and 
its running, Wilson and 't Hooft loops, anomalous 
dimensions of the quark superfields, quartic coupling, its running and 
anomalies 
(continuous and discrete). Based 
on those string theory predictions, we provided a 
large number of matchings between field theory expectations and string 
theory  results.
Interestingly, our solutions seem to realize the Fradkin-Shenker result 
for the non-existence of phase transitions when moving between higgsed and 
confining phases of a theory with fundamental fields 
\cite{Fradkin:1978dv}. 

This work leaves some open questions for future research. Indeed, it 
should be nice, among other things, to understand if the same `phase 
structure' occurs in 
type {\bf N} backgrounds in eq.(\ref{nonabmetric424}) and the behavior of 
the quartic coupling in that case. The picture of Seiberg duality for 
type {\bf N} 
backgrounds could be understood on a similar way as we did in this paper. 
The case of massive flavors, either for type {\bf A} or {\bf N} 
backgrounds also presents interest because it gives a resolution 
of the IR 
(good) singularity and its many applications. These and other topics will 
be dealt with in \cite{cnp3}. 

It should also be interesting 
to apply our treatment for quartic couplings to cascading theories.
Sorting out other dynamical aspects of 
the backgrounds presented here, like for example, the spectrum of 
glueballs, mesons and baryons \cite{Caceres:2005yx} is of main interest.
The Veneziano expansion we are adopting indicates that there 
will be interactions leading to decays between the first two types of 
excitations and suggests ways to study baryons as solitons in the BI-WZ action.

It should also be interesting to extend all this story to the type II A 
case, using backgrounds like the ones in \cite{Atiyah:2000zz}. Extensions 
to lower dimensional field theories may prove interesting for 
condensed matter problems. Finally, it should be nice to think about 
possible high energy phenomenology applications of this line of 
research.

\section*{Acknowledgments:}
In the last year or so, discussions with different 
colleagues have shaped our understanding 
and the presentation of the topics in this paper. For those discussions, 
we would like 
to thank: Gert Aarts, Riccardo Argurio,
Adi Armoni,  Francesco Benini, Francesco Bigazzi,
Aldo Cotrone,
Felipe Canoura,
Stefano Cremonesi,
Justin Foley, Seba Franco, Simon 
Hands, Prem Kumar, Biagio Lucini, Asad Naqvi, 
Alfonso V. Ramallo and
Diego Rodriguez G\'omez.
RC and AP are supported by European Commission Marie Curie Postdoctoral
Fellowships under contracts MEIF-CT-2005-024710 and MEIF-CT-2005-023373.
RC and AP's work was also partially supported by
INTAS grant, 03-51-6346, RTN contracts MRTN-CT-2004-005104 and
MRTN-CT-2004-503369, CNRS PICS \#~2530,  3059 and 3747,
 and by a European Union Excellence Grant,
MEXT-CT-2003-509661.

\appendix

\section*{APPENDIX}
\addcontentsline{toc}{section}{APPENDICES:}

\section{Beta function for $\kappa_m$}
\label{appendix2}
\setcounter{equation}{0}
The computation of the beta function for the quartic coupling in 
the magnetic case seems to be tricky, so, here we will write 
the results we get and our interpretation.

If we Seiberg dualize the coupling $\kappa_e$ in eq.(\ref{quarticelect}) we get the 
magnetic 
version $\kappa_m=1+\frac{H}{G}$. We compute the beta function from string theory 
using the radius-energy relation eq.(\ref{re}) and we obtain 
\beq
\frac{1}{\kappa_m}\beta_{\tilde{\kappa}_m}
= \frac{\partial \log[\tilde{\kappa}_m]}{\partial 
\log(\frac{\mu}{\Lambda})} = 
\frac{3}{2}\frac{\partial }{\partial\rho} 
\log\Big(1+\frac{H}{G}\Big) \approx 
-\frac{3}{4}\frac{(N_f-n_c)}{(N_f-2n_c)\rho^2}.
\label{betakappamagn}
\eeq
Now, let us look at the QFT results.
The superpotential after Seiberg duality is given by,
\beq
{\cal{W}}_m=\hat{\kappa}M M + \hat{z}\tilde{q}M q
\label{supermag}
\eeq
The classical dimensions for fields and couplings are,
\beq
[q]=m,\;\; [M]= m,\;\;[\hat{\kappa}]= m, \;\; [\hat{z}]=1
\eeq
where we used that the ``meson'' $M$ is a fundamental field in the 
magnetic theory and not a composite of electric quarks. The R-charges of 
fields and couplings, obtained from the main part of the paper 
eqs.(\ref{valuescharges2}) \footnote{For the 
values 
of R-charges in eqs.(\ref{valuescargas})-(\ref{valuescharges}) we cannot 
find an expression for the anomalous dimension of the meson superfield. 
We are not at a conformal point and the relation 
(\ref{conformalpointrelation}) should not be 
used.} are, 
\beq
R[q]=(\frac{3}{2}-2\frac{n_c}{N_f}),\;\; R[M]=( 4\frac{n_c}{N_f}-1),\;\; 
R[\hat{\kappa}]=4 (1-2 \frac{n_c}{N_f})
\eeq
and the anomalous dimensions from the main part of the paper (for the 
meson it is computed assuming that the relation between dimension and 
R-charge is the one of a fixed point), 
\beq
\gamma_q=\frac{12}{N_f}Y -\frac{1}{2} -3\frac{n_c}{N_f}\approx 
\frac{5}{2}-6\frac{n_c}{N_f}- \frac{3}{8}
\frac{(N_f-n_c)}{(N_f-2n_c)\rho^2},\;\;\; \gamma_M=-5 + 12 \frac{n_c}{N_f}
\label{anomamag}
\eeq
We can now compute
\beq
\beta_{\hat{\kappa}}= \hat{\kappa}[-1 + \gamma_M]= \hat{\kappa}(
12\frac{n_c}{N_f} -6),\;\;\; \beta_{\hat{z}}= 
\hat{z}[\gamma_q+\frac{1}{2}\gamma_M]=-\frac{3}{8}\hat{z} 
\Big(\frac{N_f-n_c}{(N_f-2n_c)\rho^2}\Big)
\label{betazkappam}
\eeq
On the other hand, we could have considered the superpotential after 
integration out of the massive meson in eq.(\ref{supermag})
\beq
{\cal{W}}_m= \tilde{\delta}\tilde{q}q\tilde{q}q
\eeq
where the double trace term is present but has no relevance here. In this 
case the beta function from QFT has the expression,
\beq
\beta_{\tilde{\delta}}=\tilde{\delta}[1+2\gamma_q]= 
(6-12\frac{n_c}{N_f}) + \frac{3}{4}\frac{(N_f-n_c)}{(N_f-2n_c)\rho^2}
\label{betamag2}
\eeq
It seems that the string theory computation in eq.(\ref{betakappamagn}) is 
reproducing the beta 
function of 
a combination of field theory couplings in eq.(\ref{betazkappam}) and 
eq.(\ref{betamag2}). Indeed, the R-symmetry neutral combination that makes 
this work is
$ \kappa_m=(\delta \hat{\kappa})^{-1} .$

\end{document}